\documentclass[aps,prx,twocolumn,nopacs,superscriptaddress,longbibliography]{revtex4-1}

\usepackage[breaklinks=true,colorlinks,citecolor=blue,linkcolor=blue,urlcolor=blue]{hyperref}
\usepackage{epsfig,mathrsfs,color,latexsym,subfigure,marginnote,graphicx,verbatim,relsize,mathrsfs,color,array,amsmath,amsfonts,amssymb,graphicx}
\begin{document}

\title{Exposing nontrivial flat bands and superconducting pairing in infinite-layer nickelates}

\author{Ruiqi~Zhang}
\affiliation{Department of Physics and Engineering Physics, Tulane University, New Orleans, LA 70118, USA}

\author{Cheng-Yi~Huang}
\affiliation{Department of Physics, Northeastern University, Boston, MA 02115, USA}

\author{Mehdi Kargarian}
\affiliation{Department of Physics, Sharif University of Technology, Tehran, Iran}

\author{Rahul Verma}
\affiliation{Department of Condensed Matter Physics and Materials Science, Tata Institute of Fundamental Research, Mumbai 400005, India}

\author{Robert S. Markiewicz}
\affiliation{Department of Physics, Northeastern University, Boston, MA 02115, USA}

\author{Arun Bansil}
\affiliation{Department of Physics, Northeastern University, Boston, MA 02115, USA}

\author{Jianwei Sun}
\email[Corresponding author:~]{jsun@tulane.edu}
\affiliation{Department of Physics and Engineering Physics, Tulane University, New Orleans, LA 70118, USA}

\author{Bahadur Singh}
\email[Corresponding author:~]{bahadur.singh@tifr.res.in}
\affiliation{Department of Condensed Matter Physics and Materials Science, Tata Institute of Fundamental Research, Mumbai 400005, India}

\begin{abstract}
Flat bands coupled with magnetism and topological orders near or at the Fermi level are well known to drive exotic correlation physics and unconventional superconductivity. Here, based on first-principles modeling combined with an in-depth symmetry analysis, we reveal the presence of topological flat bands involving low-energy Ni-$3d_{z^2}$ states in the recently discovered superconductor LaNiO$_{2}$. Our analysis demonstrates that LaNiO$_2$ is an Axion insulator with $\mathbb{Z}_{4} = 2$ and that it supports topological crystalline insulating states protected by the glide mirror symmetries. The topological flat bands in LaNiO$_{2}$ are also shown to host odd-parity superconductivity. Our study indicates that the nickelates would provide an interesting materials platform for exploring the interplay of flat bands, topological states, and superconductivity. 
\end{abstract} 

\maketitle

{\it Introduction.}
Flat bands in quantum materials provide a robust basis for investigating many-body physics, including non-Fermi liquids and unconventional superconductivity interwoven with nontrivial electronic states~\cite{Regnault2021,Ma2020,Yin2022,Wen2011,SunKai2011,Cao2018}. Suppression of electronic kinetic energy in a flat band produces electron localization in real space and the emergence of highly degenerate electronic states. Although models and materials have been proposed~\cite{Wen2011,SunKai2011}, the experimental realization of flat bands has been limited mainly to moiré lattice materials such as twisted bilayer graphene (TBLG) and kagome materials~\cite{Cao2018,Yin2022}. The mechanism in moiré materials  involves the flattening of the Dirac cone dispersion through spatial variations of the interlayer coupling that restructures the underlying lattice and yields a smaller Brillouin zone~\cite{Bistritzer2011}. The moiré  materials although exhibit nontrivial states coexisting with superconducting, magnetic, and other correlated states, but controlling their properties requires complex tuning of structural parameters.  For this reason, stoichiometric crystals with flat bands are more appealing for exploring the physics of high electron density coupled with nontrivial topology~\cite{Regnault2021,Ma2020,Yin2022}, and the kagome materials are drawing intense current interest~\cite{Yin2022,Chen2021,Kang2022,Jiang2023,Singh2023,Wang2023,Kang2020,Singh2023}. However, the flat bands in the kagome materials usually lie away from the Fermi level~\cite{Yin2022}, and therefore, it is not clear how these bands participate in driving superconductivity. It is important therefore to find superconducting materials that support low-energy flat bands that coexist with other correlated states~\cite{Regnault2021,Ma2020,Yin2022}.  

The cuprates have been shown recently to host high-order van Hove singularities (VHSs) and flat bands which could drive correlated phases such as superconductivity and supermetals~\cite{Markiewicz2023}. Here we characterize the topological flat bands and superconducting pairing symmetries of the recently discovered infinite-layer (IL) nickelate superconductors that have emerged as a promising analog of the cuprates~\cite{Li2019a, Osada2020a, ZhangS2021, Osada2021, Sawatzky2019, Norman2020, Pickett2021}. Extensive experimental and theoretical efforts have been devoted to finding similarities and differences between the infinite-layer (IL) nickelates and the cuprates ~\cite{Hepting2020,Liu2020, Choi2020,Jiang2019,Lechermann2020,Leonov2020,Lee2004,Zhang2020d,Sakakibara2019,Botana2020,Karp2020,Goodge2021,Jiang2020,Zhang2021,2022arXiv220700184Z,Lane2023}. For instance, competing charge orders~\cite{2021arXiv211202484R,Krieger2021,Tam2021}, superconductivity~\cite{2021arXiv211202484R, OsadaAM2021}, and magnetism~\cite{Fowlie2022,Lu2021science} has been reported recently in the $undoped$ IL-nickelates~\cite{Li2019a,2021arXiv211113668O, Lu2021science}. These findings sharply contrast with the $undoped$ cuprates, which host a well-defined AFM ground state. The pairing symmetry in IL-nickelates, however, remains unclear~\cite{Sakakibara2019,Gu2020, Kitatani2020,2022arXiv220110038C}. Although the participation of Ni 3$d_{z^2}$-like flat bands in superconducting pairing is indicated~\cite{Kreisel2022PRL}, their topological character and the associated pairing symmetries have not been identified~\cite{GaoNSR2020}. 

In this Letter, we unveil the presence of topological flat bands in the parent nickelate superconductor LaNiO$_2$ and discuss the associated pairing symmetries. LaNiO$_2$ has been shown to support various low-energy competing magnetic phases with $C$-type antiferromagnetic ($C-$AFM) ground state~\cite{Zhang2021,2022arXiv220700184Z}. The $C-$AFM state supports flat bands primarily composed of Ni $3d_{z^2}$ states with VHSs at the Fermi level, which would enhance many-body interactions and drive exotic correlated phenomena~\cite{Zhang2021,ChoiPRR2020}. We delineate the topological character of LaNiO$_2$ and its nontrivial topological state.  Our in-depth symmetry analysis of the bulk bands and surface states shows that the $C-$AFM phase realizes an axion insulator state with $\mathbb{Z}_{4} = 2$ and a glide-mirror-symmetry protected topological crystalline insulating (TCI) state. We also characterize the superconducting pairing symmetries associated with the bulk flat bands at the Fermi level and discuss the possibility of an odd-parity pairing in LaNiO$_2$. Our study indicates that LaNiO$_2$ is a materials promising platform for exploring the interplay of topological flat bands and unconventional superconductivity.

{\it Methods.}
All calculations were performed by using the projector-augmented wave method~\cite{Kresse1999} as implemented in the Vienna {\it ab initio} simulation package (VASP)~\cite{Kresse1993, Kresse1996}. A high-energy cutoff of 520 eV was used to truncate the plane-wave basis set. The exchange-correlation effects were treated using the strongly constrained and appropriately normed (SCAN) density functional without any adjustable parameters~\cite{Sun2015}. The crystal structures and ionic positions were fully optimized using a force convergence criterion of 0.01  eV/\AA{} for each atom along with a total energy tolerance of 10$ ^{-6} $ eV. Notably, substantial recent studies have shown the efficacy of SCAN for accurately modeling ground states and electronic structure of strongly correlated systems such as the cuprates~\cite{Zhang2020, Furness2018, Lane2018} and nickelates~\cite{Zhang2021,2022arXiv220700184Z}, binary $3d$ oxides~\cite{Zhang2020b}, and especially the \textit{f}-electron system SmB$_{6}$~\cite{ZhangR2020} due to the reduction of self-interaction error in comparison with other widely used density functionals~\cite{Zhang2020b}. We adopted a 16 $\times$ 16 $\times$ 16 $\Gamma$-centered $k$ mesh to sample the primitive bulk Brillouin zone (BZ) and total energy calculations. Spin-orbit coupling (SOC) effects were included self-consistently. Topological properties were calculated by generating a material-specific tight-binding model Hamiltonian using the VASP2WANNIER90 interface~\cite{MOSTOFI20142309}. La $f$ and $d$, Ni $d$, and  O $p$ states were included for constructing the Wannier functions. The surface spectral weight of semi-infinite slabs was calculated using an iterative Green's function method~\cite{GreenF1983,sancho1984quick}.

\begin{figure}[t!]
\centering
\includegraphics[width=0.99\columnwidth]{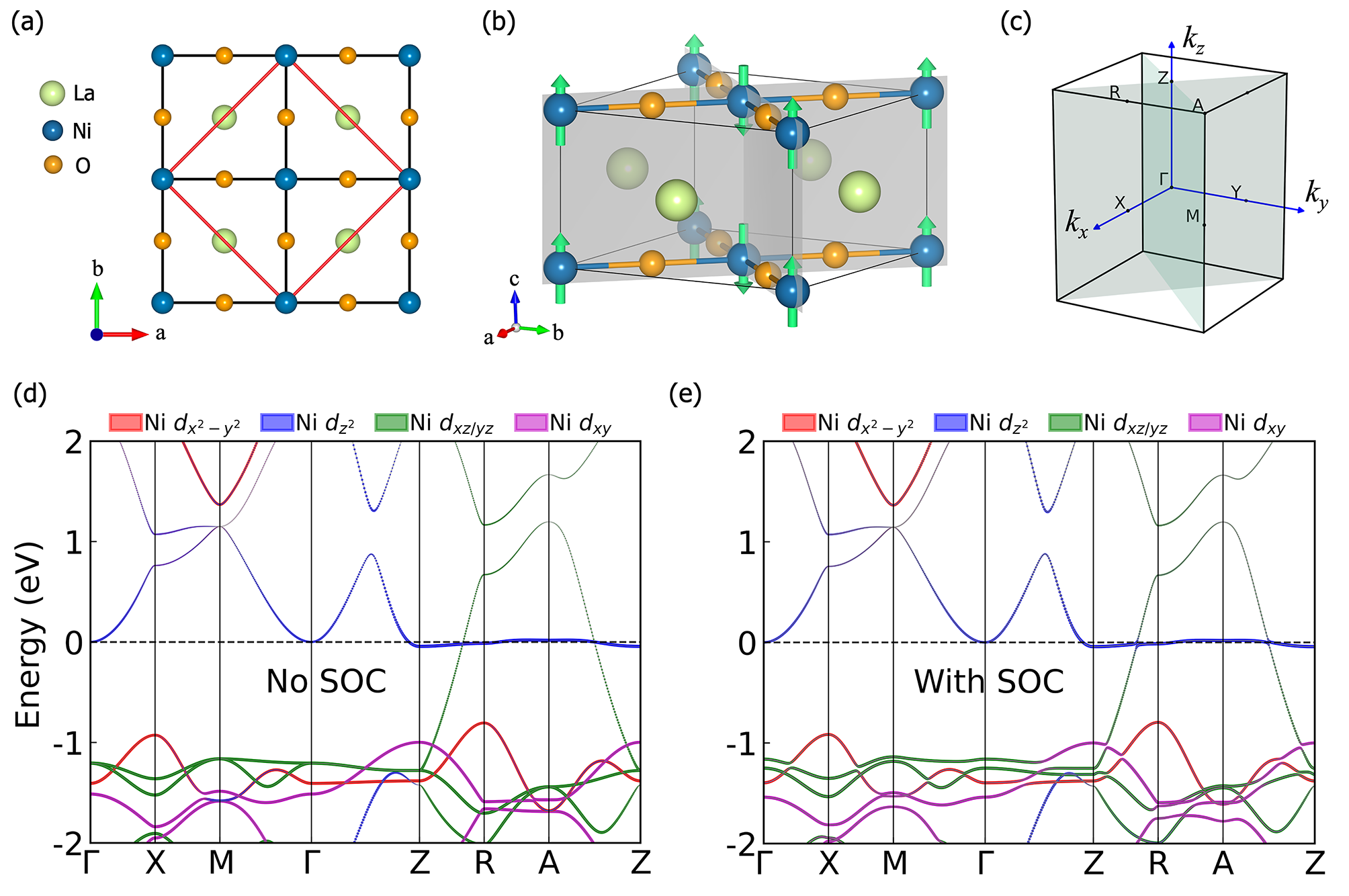}
\caption{ (a) A $2\times2$ representation of parent nonmagnetic structure of LaNiO$_{2}$ with the $C-$AFM unit cell marked in red color. Green, blue, and yellow balls represent La, Ni, and O atoms, respectively. (b) The perspective view of the $C-$AFM phase of LaNiO$_{2}$, where $\mathcal{M}_{110}$ and $\mathcal{M}_{1\bar{1}0}$ planes are highlighted. The green arrows denote magnetic moment directions. (c) The first Brillouin zone (BZ) of $C-$AFM with relevant high-symmetry points. The $\mathcal{M}_{110}$ and $\mathcal{M}_{1\bar{1}0}$ planes are also shown. (d)-(e) Calculated bulk band structure of the $C-$AFM phase of LaNiO$_{2}$ (d) without and (e) with spin-orbit coupling. The orbital decompositions to various states are identified.}
\label{fig:fig1}
\end{figure}

{\it Crystal structure and electronic properties.}
The parent nonmagnetic LaNiO$_{2}$ crystallizes in the tetragonal Bravais lattice with space group $P4/mmm$ ($\#123, D_{4h}$). The La, Ni, and O atoms occupy Wyckoff positions $1d:~\{\frac{1}{2}, \frac{1}{2}, \frac{1}{2}\}$, $1a:~\{0, 0, 0\}$, and $2f:~\{\frac{1}{2}, 0, 0\}$, respectively, in the lattice. The magnetic unit cell of $C-$AFM phase is constructed considering a $\sqrt{2}\times\sqrt{2}\times1$ supercell of nonmagnetic LaNiO$_{2}$ (see Fig.~\ref{fig:fig1}(a)). In the $C-$AFM phase, the magnetic moments of Ni atoms are aligned parallel to the tetragonal $c$ axis as illustrated in Fig.~\ref{fig:fig1}(b). Particularly, the magnetic interaction of Ni atoms is antiferromagnetic in the plane whereas it is ferromagnetic in the out-of-plane direction. The calculated magnetic moment of Ni is $\sim 1~\mu_{B}$, which is consistent with previous DFT+$U$~\cite{ChoiPRR2020} and DMFT calculations~\cite{Leonov2020}. Note that despite the absence of long-range orders in LaNiO$_2$ even at low temperatures, experiments reported the presence of local magnetic moments on Ni atoms and strong AFM interactions with competing charge orders in nickelates~\cite{Fowlie2022,Lu2021science}. In this context, our recent study has discovered a variety of low-energy competing magnetic phases of LaNiO$_2$ with $C-$AFM phase as the lowest energy state~\cite{Zhang2021,2022arXiv220700184Z}. We thus focus on the symmetry properties and topological properties of the $C-$AFM phase. It retains the $D_{4h}$ point-group symmetry of the parent phase when the SOC effects are included. It contains 16 symmetry operations including an inversion $\mathcal{I}$ symmetry and glide-mirror $\mathcal{G}_{v}=\{\mathcal{M}_{v}| \frac{1}{2}\frac{1}{2}0\}$ symmetries highligted in Fig.~\ref{fig:fig1}(b). Figure~\ref{fig:fig1}(c) shows the first BZ of the $C-$AFM phase with various high-symmetry points. 

The bulk band structure of the $C-$AFM phase without SOC is shown in Fig.~\ref{fig:fig1}(d). A pronounced antiferromagnetic splitting of $\sim$2eV in Ni $3d_{x^2-y^2}$ bands is evident in accord with earlier DFT+$U$~\cite{ChoiPRR2020} and experimental studies~\cite{Tam2021}. More importantly, we observe the presence of a nearly flat band pinned precisely at the Fermi level on the $k_{z} = \pi/c$ plane. This flat band intersects with another itinerant band to form nodal band crossings along the $Z-R$ and $A-Z$ high-symmetry directions. The detailed orbital analysis shows that the flat band is predominantly composed of Ni 3$d_{z^2}$ orbitals and crosses with O 2$p$ and Ni 3$d_{xz/yz}$ derived itinerant band to form a nodal band crossing. Upon considering the SOC effects in Fig.~\ref{fig:fig1}(e), a minuscule gap opens up these crossing points, forming a continuous local band gap between the valence and conduction bands.  

\begin{figure}[t!]
\centering
\includegraphics[width=0.99\columnwidth]{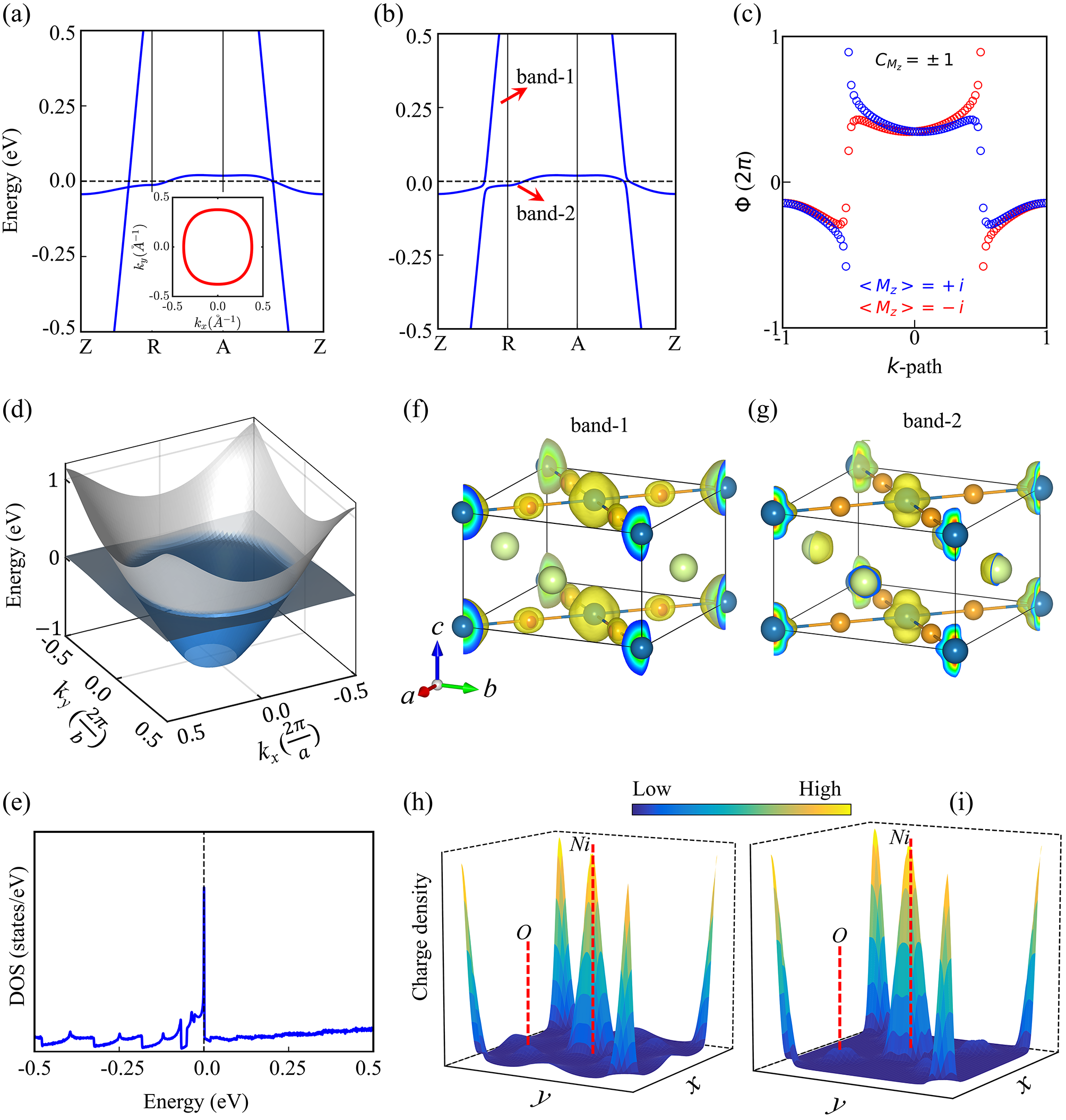}
\caption{Closeup of the bulk band structure of $C-$AFM phase (a) without and (b) with spin-orbit coupling. Band-1 and band-2 are marked in (b) to facilitate the discussion (see text for details).  (c) The evolution of the mirror winding number on the $k_x = k_y$ plane. The non-zero mirror Chen number shows the TCI phase in the $C-$AFM phase of LaNiO$_2$. (d) $E-k_x-k_y$ plot resolving the crossings band structure on the $k_z = \pi/c$ plane. (e) Calculated density of states (DOS). A large peaked DOS at the Fermi level highlights VHSs due the the flat band energy. (f)-(g) The charge density associated with band-1 and band-2 represented in the crystal lattice.  (h)-(i) The charge density associated with band-1 and band-2 represented in the $k_z = \pi/c$ plane. A power decay of the Bloch wavefunctions associated with Ni atoms is revealed.}
\label{fig:fig2}
\end{figure}

{\it Flat band and the nontrivial topological state.}
To characterize the flat band and nontrivial state, we present the close-up of the band structure without and with SOC in Figs.~\ref{fig:fig2}(a) and \ref{fig:fig2}(b), respectively. Interestingly, a scan of the nodal crossings in the full BZ unfolds a nearly dispersionless nodal line on the $k_{z} = \pi/c$ plane without SOC as illustrated in the inset of Fig.~\ref{fig:fig2}(a). The inclusion of SOC drives an inverted insulating gap of $\leq 5$ meV between band-1 and band-2 in Fig.~\ref{fig:fig2}(b). For clarity, we denote the two intersecting bands at the Fermi level as band-1 and band-2 in Fig.~\ref{fig:fig2}(b). Band-1 primarily arises due to strong hybridization between the Ni 3$d$ and O 2$p$ orbitals (Fig.~\ref{fig:fig2}(f)), whereas band-2 exhibits Ni 3$d_{z^2}$ and La 5$d$ states as shown in Fig.~\ref{fig:fig2}(g).  These findings not only infer a nontrivial character of the $C-$AFM phase but also underscore the necessity of a multiband model to accurately describe the low-energy physics in nickelates. We present a three-dimensional ($E-k_x-k_y$) rendition of the band structure near the Fermi level in Fig.~\ref{fig:fig2}(d). It highlights the nearly isotropic nature of the flat bands at the $k_{z} = \pi/c$ plane. Such a flat band drives high-order Van Hove singularities (VHSs) in the density of states (Fig.~\ref{fig:fig2}(e)) similar to $high-T_c$ cuprates superconductors~\cite{Markiewicz2023}. The presence of flat bands and VHSs at the Fermi level can introduce competition among various lattice-distorted and correlated electronic states in nickelates as found in experiments. To determine the origin of the flat bands, we plot the charge density of band-1 and band-2 in Figs.~\ref{fig:fig2}(f)-\ref{fig:fig2}(g), and \ref{fig:fig2}(h)-\ref{fig:fig2}(i). We find a power-law decay of Ni wavefunctions on the $k_{z} = \pi/c$ plane, which results in highly restricted in-plane hopping between Ni spins. In contrast, the out-of-plane Ni wavefunctions are less localized, giving rise to bands with notable dispersion. This scenario aligns with Hund's coupling rules, signifying that in-plane hopping terms are significantly suppressed due to the conservation of spin while hopping along the $c$-axis is free from such a constraint.

The presence of an inverted band gap between band-1 and band-2 in the $C-$AFM phase of LaNiO$_{2}$ hints towards a topological nontrivial state. To reveal the nontrivial band character, we compute the parity-based higher-order  $\mathbb{Z}_{4}$ invariant which is given as 
\begin{equation}
\mathbb{Z}_{4} = \sum_{i= 1}^{8}\sum_{n= 1}^{n_{occ}} \frac{1+\xi_n(\Lambda_i)}{2}~ \mathrm{mod ~ 4}
\end{equation}
where $\xi_n$ is the parity eigenvalue ($\xi_n = \pm1$) of $n^{th}$ band at time-reversal invariant momentum point $\Lambda_i$. The $\mathbb{Z}_{4}$ is well defined even for a magnetic material preserving inversion symmetry. A centrosymmetric insulator with $\mathbb{Z}_{4}=2$ indicates an axion insulator state with a quantized topological magnetoelectric effect with axion coupling $\theta=\pi$~\cite{Yuanfeng2019}. Based on the parity eigenvalues of occupied states, we obtained $\mathbb{Z}_{4}$ of 2, which demonstrates that the $C-$AFM phase of LaNiO$_{2}$ is an antiferromagnetic axion insulator. It can support the half-quantized quantum Hall effect if the associated surface states are gapped. 

\begin{figure}[htpb]
\centering
\includegraphics[width=0.99\columnwidth]{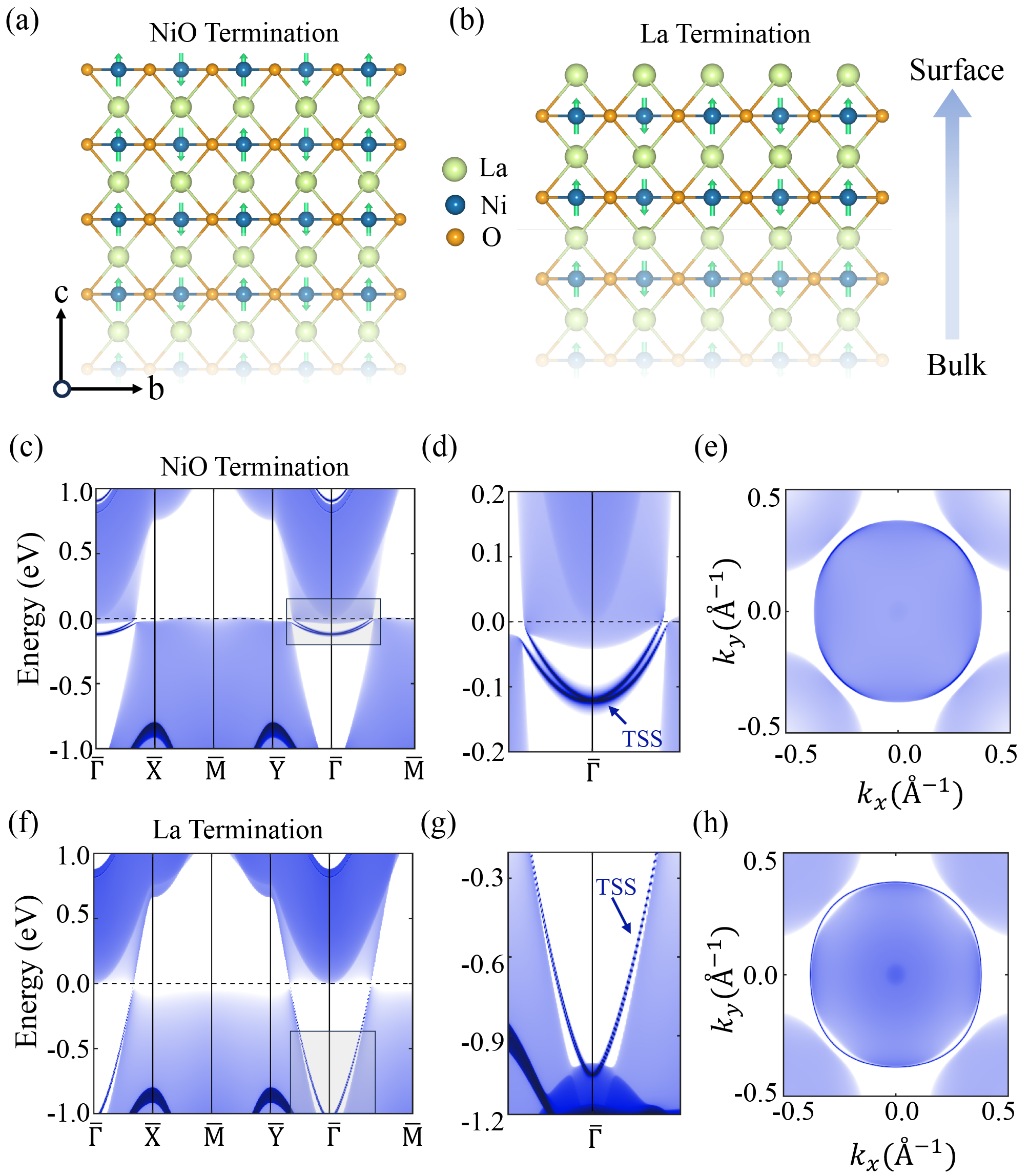}
\caption{Lattice structure of a semi-infinite slab of the $C-$AFM phase of LaNiO$_2$ with (a) NiO and (b) La terminations. (c)-(e) The calculated (001) surface band structure for NiO surface termination (c) with a closeup of surface bands in (d) and associated Fermi band counter in (e). The dark blue color marks the surface states. The horizontal dashed line marks the Fermi level. (f)-(h) Same as (c)-(e) but for La surface termination.}
\label{fig:fig3}
\end{figure}

Since the system respects $\mathcal{G}_{v_1}=\{\mathcal{M}_{v_1}| \frac{1}{2}\frac{1}{2}0\}$, we also evaluated the mirror Chern number on the $k_{x}=k_{y}$ plane. The evolution of the Wannier charge centers (WCCs) associated with two mirror eigen-sectors is plotted in Fig.~\ref{fig:fig2}(c), which reveals the nontrivial binding of WCCs and a mirror Chern number of -1.  The nontrivial mirror Chern number guarantees the existence of nontrivial Dirac cone states on the surface respecting glide-mirror $\mathcal{G}_{v_1}=\{\mathcal{M}_{v_1}| \frac{1}{2}\frac{1}{2}0\}$ symmetry. We present the (001) surface band spectrum for Ni-O and La terminated surfaces in Fig.~\ref{fig:fig3}. In Figs.~\ref{fig:fig3}(a)-\ref{fig:fig3}(b), we illustrate the two possible (001) surface terminations that expose the Ni-O and La atomic planes in the surface layer, respectively. The surface electronic structures associated with the Ni-O and La terminated surfaces are shown in Figs.~\ref{fig:fig3}(c)-(e) and \ref{fig:fig3}(f)-\ref{fig:fig3}(h), respectively. The clear surface Dirac cone states are seen crossings at the $\bar{\Gamma}$ point on both the terminations. However, these surface states show distinct energy-momentum dispersion on different terminated surfaces. Specifically, the surface states on the Ni-O surface show a nearly flat energy dispersion with a bandwidth of $\sim$0.1eV (Figs.~\ref{fig:fig3}(c)-\ref{fig:fig3}(d)). The surface states overlap with the projected flat bulk bands for the Ni-O surface at the Fermi level in full surface BZ as revealed in the Fermi band contours in Fig.~\ref{fig:fig3}(e). In contrast, the surface states on the La terminated surface have a larger bandwidth with a higher group velocity (Figs.~\ref{fig:fig3}(f)-\ref{fig:fig3}(g)). The associated Fermi band contour in Fig.~\ref{fig:fig3}(h) is separated from bulk bands in momentum space. The reason for such a difference in energy dispersion on two terminated surfaces can be attributed to a suppressed hopping strength on Ni-O termination due to anti-parallel magnetic moments between Ni atoms while there is no such restriction for non-magnetic La termination on the surface. 

We turn now to discuss the symmetry protection of the observed surface Dirac fermions. The glide mirror plane $\mathcal{G}_{v_1}: (x,y,z)\rightarrow \left(y+\frac{1}{2}, x+\frac{1}{2}, z \right)$ reflects the states in momentum space at $(k_x,k_y,k_z)$  to $(k_y,k_x,k_z)$. The states on the $(k_x,k_x,k_z)$ plane can be labeled by glide eigenvalues. To obtain these glide eigenvalues, the relation $\hat{\mathcal{G}}_{v_1}^2=\{-E|110\}$ holds for a spin-full system with action on the Bloch states as $\hat{\mathcal{G}}_{v_1}^2|\mathbf{k}\rangle=-e^{-i(k_x+k_y)}|\mathbf{k}\rangle$. Therefore, the glide eigenvalues on the $(k_x,k_x,k_z)$ plane read as $g_{\pm}=\pm i e^{-ik_x}$. In particular, on the line $k_x=k_y=0$, the glide eigenvalues become pure imaginary, $g_{\pm}=\pm i$, and the signs are swapped on the line $k_x=k_y=\pi$, {\it i.e.}, $g_{\pm}=\mp i$. There could be thus unavoided band crossing on the surface preserving the glide symmetry ($\bar{\Gamma}-\bar{M}$ line on (001) surface) as found at $\bar{\Gamma}$ point in Fig.~\ref{fig:fig3}.  

Similarly, another vertical glide mirror plane $\mathcal{G}_{v_2}: (x,y,z)\rightarrow \left(x+\frac{1}{2}, -y+\frac{1}{2}, z \right)$ dictates that the states at $(k_x,k_y,k_z)$ reflects to  $(k_x,-k_y,k_z)$. The eigenstates on $k_y=\{0,~\pi\}$ planes can be thus labeled by $\mathcal{G}_{v_2}$ glide eigenvalues. Using the relations $\hat{\mathcal{G}}_{v_2}^2=\{-E|110\}, ~~~ \hat{\mathcal{G}}_{v_2}^2|\mathbf{k}\rangle=-e^{-i(k_x+k_y)}|\mathbf{k}\rangle$, we obtain $g_{\pm}(k_x, k_y=0)=\pm i e^{-ik_x/2},~~~~g_{\pm}(k_x, k_y=\pi)=\pm e^{-ik_x/2}$ on the $k_y=\{0,~\pi\}$ planes. On the glide symmetric plane $k_y=0$ and for the subspace restricted to the $k_z$ axis ({\it i.e.}, $k_x=0$), the glide eigenvalues become $g_{\pm}=\pm i$. This dictates the band degeneracy at $\bar{\Gamma}$ and thus, the surface states can be gapless along $\bar{\Gamma}-\bar{X}$ as well. These symmetry properties indicate that the surface Dirac fermions are protected by glide mirror symmetry, and thus, the $C-$AFM phase of LaNiO$_{2}$ realizes a symmetry-protected topological crystalline insulator state.

{\it Possible nontrivial superconducting states.}
The preceding analysis demonstrates that LaNiO$_{2}$ is a topological crystalline insulator with nontrivial surface Dirac Fermion states. These states could become superconducting through the bulk proximity effect. We now analyze the possibility of nontrivial superconducting pairing in the bulk states of LaNiO$_{2}$. Under the glide operation, the symmetric kinetic bulk Hamiltonian transforms as $\hat{\mathcal{G}}(\mathbf{k}) H(\mathbf{k})\hat{\mathcal{G}}(\mathbf{k})^{\dagger}=H(\mathcal{M}_{v}\mathbf{k})$, where $\hat{\mathcal{G}}(\mathbf{k})~(\equiv \{\mathcal{M}_{v_1/v_2}|\frac{1}{2}\frac{1}{2}0\})$ represents the glide mirror operator in momentum space acting on spin, orbital, and sublattice degrees of freedom.  The superconducting model with gap function $\Delta(\mathbf{k})$ is described by the BdG Hamiltonian,
\begin{equation}
H_{\mathrm{BdG}}(\mathbf{k}) = 
 \begin{pmatrix}
  H(\mathbf{k}) & \Delta(\mathbf{k}) \\
 \Delta^{\dagger}(\mathbf{k}) & -H^{\mathrm{T}}(-\mathbf{k})
 \end{pmatrix}.
\end{equation}

The bulk symmetries imply that $\hat{\mathcal{G}}_{\mathrm{BdG}}(\mathbf{k}) H_{\mathrm{BdG}}(\mathbf{k})\hat{\mathcal{G}}_{\mathrm{BdG}}(\mathbf{k})^{\dagger}=H_{\mathrm{BdG}}(\mathcal{M}_{v}\mathbf{k})$. Here the symmetry operations are extended to the particle-hole space as $\hat{\mathcal{G}}_{\mathrm{BdG}}(\mathbf{k})=\hat{\mathcal{G}}(\mathbf{k})\oplus [\hat{\mathcal{G}}^{\pm}(-\mathbf{k})]^*$. To satisfy this symmetry constraint, the gap function should transform as $\hat{\mathcal{G}}(\mathbf{k}) \Delta(\mathbf{k})\hat{\mathcal{G}}^{\mathrm{T}}(-\mathbf{k})=\pm \Delta(\mathcal{M}_{v}\mathbf{k})$. We define $\hat{\mathcal{G}}^{\pm}(-\mathbf{k})=\pm \hat{\mathcal{G}}(-\mathbf{k})$ depending on the even (+) or odd (-) transformation of the gap function under the symmetry operation. For odd-symmetric gap functions, the lattice symmetry operation and particle-hole transformation, $\hat{\mathcal{C}}H_{\mathrm{BdG}}(\mathbf{k})\hat{\mathcal{C}}^{-1}=-H_{\mathrm{BdG}}(-\mathbf{k})$ should anticommute {\it i.e.} $\{\hat{\mathcal{C}}, \hat{\mathcal{G}}\}=0$. 
These symmetry constraints restrict the Hilbert space of Bloch states to one-dimensional lines $k_x=k_y=0$ and $k_x=k_y=\pi$, which corresponds to $\Gamma-Z$ and $M-A$ symmetry directions in the bulk BZ, respectively. The Bloch states on these symmetry lines thus acquire pure imaginary glide eigenvalues $g_{\pm}=\pm i$: $\mathcal{H}=\bigoplus_{k_z}\left( \mathcal{H}_{k_z}^{+i}\oplus\mathcal{H}_{k_z}^{-i}\right)$. 

Let us consider superconducting states with odd-symmetric gap functions. Each eigenvalue sector is preserved under the action of particle-hole symmetry operation due to the anticommutation relation $\{\hat{\mathcal{C}}, \hat{\mathcal{G}}\}=0$, {\it i.e.}, a glide eigenstate with definite eigenvalue is preserved upon the action of $\hat{\mathcal{C}}$ \cite{Nogaki:PRR2021}. Therefore, each sector is a one-dimensional superconductor that breaks the time-reversal symmetry. It belongs to the $D$ class of AZ classification. The superconducting states in this class are characterized by $\mathbb{Z}_2$ invariant defined as 
\begin{equation}
\nu^{\pm}=\frac{1}{2\pi}\oint A^{\pm}(k)dk,
\end{equation}            
where $A^{\pm}(k)=i\sum_{n}\langle\psi^{\pm}_n(k)|\partial_{k}\psi^{\pm}_n(k)\rangle$ is the sum of the Berry connection of occupied BdG states with a definite glide eigenvalue, and the integration is over the loops $Z - \Gamma - Z$ or $A - M - A$. For an odd-parity superconductor, the topological invariant is $\mathbb{Z}_2$ nontrivial and it is shown that the invariant can be deduced from the structure of the Fermi surface of the normal state. The invariant is related to the band structure as \cite{FuB2010, Sato:prb2010}:
 
\begin{equation}\label{invariant}
(-1)^{\nu^{\pm}}=\prod_{n}\mathrm{sgn}\varepsilon^{\pm}_{n}(\Lambda_i)\mathrm{sgn}\varepsilon^{\pm}_{n}(\Lambda_j),
\end{equation}
where $\varepsilon^{\pm}_{n}(\Lambda_i)$ is the eigenstate of the normal Hamiltonian $H(\mathbf{k})$ at $\Lambda_i=\Gamma (M)$ and $\Lambda_j=Z (A)$. From \eqref{invariant}, the invariant is defined modulo 2, and therefore, it can be written as
$
(-1)^{\nu^{\pm}}=(-1)^{p_0} ~~~\Rightarrow~~~ \nu^{\pm}=p_0~~\mathrm{mod}~~2,
$
where $p_0$ is the number of times the energy bands with a given glide eigenvalue cross the Fermi level between $\Lambda_i$ and $\Lambda_j$. Using the band structure shown in Figs.~\ref{fig:fig1}(e) and \ref{fig:fig2}(b), we give the sign of energy bands crossing the Fermi level in Table~\ref{Msgn}. These results show that $\mathbb{Z}_2$ invariant, $\nu^{\pm}=+1$, is nontrivial for an odd-parity superconductor. This implies that the surface perpendicular to the $z$ direction should support zero-energy modes at $\bar{\Gamma}$. 

As a possible odd-parity pairing candidate, the superconducting state could be a pair-density wave (PDW) with $B_{2u}$ representation. This irreducible representation allows for $d$-wave superconductor, $k_{x}^2-k_y^2$, within the layers while the sign of the superconducting gap alternates between the layers. A similar PDW state has been proposed as topological crystalline superconductivity in heavy-fermion CeRh$_2$As$_2$~\cite{Nogaki:PRR2021}.

\begin{table}[h]
\setlength{\belowcaptionskip}{0.5cm}
\renewcommand\arraystretch{1.4}
\caption{Sign of the mirror eigenvalues of energy bands crossing Fermi level at $\Lambda$ in Fig.~\ref{fig:fig1}(e).}
\centering
\tabcolsep 0.2cm
\begin{tabular}{ |c|cccccc| } 
\hline
$\Lambda$ & $\Gamma$ & X & M & Z & R & A   \\
\hline
$\mathrm{sgn}$[$\varepsilon(\Lambda)$] & + & + & + & - & - & + \\
\hline
\end{tabular}\label{Msgn} 
\end{table}

The symmetry of the magnetic state also allows a $\mathbb{Z}_2$ classification of possible gapped superconducting states in the layered nickelates. The magnetic state breaks both the time-reversal and translational symmetry. Nevertheless, a combination of these operations is preserved in the antiferromagnetic state \cite{Moore:PRB2010}. We denote the time-reversal operation by $\Theta$, and the half-cell translation by $T_{1/2}\equiv T_{\mathbf{a}/2+\mathbf{b}/2}$. The combined operation, $\tilde{\Theta}=\Theta T_{1/2}$, is the symmetry of the magnetic state.  For simplicity, let us define new lattice vectors $\mathbf{a}_1=\mathbf{a}+\mathbf{b}$, $\mathbf{a}_2=\mathbf{a}-\mathbf{b}$ and $\mathbf{a}_3=\mathbf{c}$. The corresponding components of wave vector reads as $k_1=\mathbf{k}\cdot\mathbf{a}_1$, $k_2=\mathbf{k}\cdot\mathbf{a}_2$ and $k_3=\mathbf{k}\cdot\mathbf{a}_3$. In momentum space, $T_{1/2}^{2}=e^{ik_1},~\tilde{\Theta}^2=-e^{ik_1}$, and $\tilde{\Theta}H_{\mathrm{BdG}}(k_1,k_2,k_3)\tilde{\Theta}^{-1}=H_{\mathrm{BdG}}(-k_1,-k_2,-k_3)$ with $\Theta^2=-1$ for spin-$1/2$ systems. On the plane $k_1=0$ ($k_x+k_y=0$)  or another $C_4$-symmetric equivalent plane, {\it i.e.}, the plane $k_2=0$ for half translation along $\mathbf{a}_2$, we have $\tilde{\Theta}^2=-1$. Thus, in analogous to time-reversal invariant insulators, the Hilbert space of Bloch states with gapped spectrum restricted to the plane $(k_2,k_3)$ can be classified by $\mathbb{Z}_2$ invariant $\nu=1$ ($\nu=0$) corresponds to topological (trivial) ground state. Note that there is no such classification on the $k_1=\pi$ plane since $\tilde{\Theta}^2=+1$. Therefore, the symmetry of the magnetic state allows for a $\mathbb{Z}_2$ classification of a fully gapped superconductor on the $(k_2,k_3)$ plane. However, whether the fully gapped superconductor on this plane is topologically nontrivial depends on the details of the band structure of $H_{\mathrm{BdG}}$, which constitutes an interesting problem for future studies. 

{\it Conclusion.--}
It is widely accepted in the cuprates that the competing inhomogeneous phases and VHSs play a key role in producing their superconducting state. This, however, is not widely discussed in the recently discovered nickelate superconductors despite their similarities with the cuprates. Our in-depth first-principles calculations and symmetry analysis reveal the existence of a flat band in the undoped nickelate superconductor LaNiO$_{2}$. This flat band mainly involves Ni 3$d_{z^2}$ states and it crosses a hybrid state composed of Ni 3$d_{xz/yz}$ and O 2$p$ orbitals to generate a band inversion in the bulk spectrum. We show that the $C-$AFM phase of LaNiO$_{2}$ realizes an axion insulator state with a higher-order invariant  $\mathbb{Z}_{4} = 2$ and mirror-symmetry-protected topological crystalline insulator state with nontrivial surface states. We also show that the superconducting state in LaNiO$_{2}$ possesses a nontrivial topological character. Our study not only highlights the similarities between the $undoped$ nickelates and the $doped$ cuprate superconductors but it also shows that the nickelates would provide an interesting materials platform for exploring the interplay of topological states and nontrivial superconductivity. 

\section*{Acknowledgements}
R.Z. and J.S. acknowledge the support of the U.S. Office of Naval Research (ONR) Grant No. N00014-22-1-2673 and it benefited from  NSF’s Advanced Cyberinfrastructure Coordination Ecosystem, and the National Energy Research Scientific Computing Center. The work at Northeastern University was supported by the National Science Foundation through NSF-ExpandQISE award \#2329067 and it benefited from Northeastern University’s Advanced Scientific Computation Center and the Discovery Cluster. The work at TIFR Mumbai was supported by the Department of Atomic Energy of the Government of India under project number 12-R\&D-TFR-5.10-0100 and benefitted from the computational resources of TIFR Mumbai. 

R. Z., C.-Y. H., and M. K. contributed equally to this work. 

\bibliography{Ref}

\begin{thebibliography}{67}%
\makeatletter
\providecommand \@ifxundefined [1]{%
 \@ifx{#1\undefined}
}%
\providecommand \@ifnum [1]{%
 \ifnum #1\expandafter \@firstoftwo
 \else \expandafter \@secondoftwo
 \fi
}%
\providecommand \@ifx [1]{%
 \ifx #1\expandafter \@firstoftwo
 \else \expandafter \@secondoftwo
 \fi
}%
\providecommand \natexlab [1]{#1}%
\providecommand \enquote  [1]{``#1''}%
\providecommand \bibnamefont  [1]{#1}%
\providecommand \bibfnamefont [1]{#1}%
\providecommand \citenamefont [1]{#1}%
\providecommand \href@noop [0]{\@secondoftwo}%
\providecommand \href [0]{\begingroup \@sanitize@url \@href}%
\providecommand \@href[1]{\@@startlink{#1}\@@href}%
\providecommand \@@href[1]{\endgroup#1\@@endlink}%
\providecommand \@sanitize@url [0]{\catcode `\\12\catcode `\$12\catcode
  `\&12\catcode `\#12\catcode `\^12\catcode `\_12\catcode `\%12\relax}%
\providecommand \@@startlink[1]{}%
\providecommand \@@endlink[0]{}%
\providecommand \url  [0]{\begingroup\@sanitize@url \@url }%
\providecommand \@url [1]{\endgroup\@href {#1}{\urlprefix }}%
\providecommand \urlprefix  [0]{URL }%
\providecommand \Eprint [0]{\href }%
\providecommand \doibase [0]{http://dx.doi.org/}%
\providecommand \selectlanguage [0]{\@gobble}%
\providecommand \bibinfo  [0]{\@secondoftwo}%
\providecommand \bibfield  [0]{\@secondoftwo}%
\providecommand \translation [1]{[#1]}%
\providecommand \BibitemOpen [0]{}%
\providecommand \bibitemStop [0]{}%
\providecommand \bibitemNoStop [0]{.\EOS\space}%
\providecommand \EOS [0]{\spacefactor3000\relax}%
\providecommand \BibitemShut  [1]{\csname bibitem#1\endcsname}%
\let\auto@bib@innerbib\@empty
\bibitem [{\citenamefont {Regnault}\ \emph {et~al.}(2022)\citenamefont
  {Regnault}, \citenamefont {Xu}, \citenamefont {Li}, \citenamefont {Ma},
  \citenamefont {Jovanovic}, \citenamefont {Yazdani}, \citenamefont {Parkin},
  \citenamefont {Felser}, \citenamefont {Schoop}, \citenamefont {Ong},
  \citenamefont {Cava}, \citenamefont {Elcoro}, \citenamefont {Song},\ and\
  \citenamefont {Bernevig}}]{Regnault2021}%
  \BibitemOpen
  \bibfield  {author} {\bibinfo {author} {\bibfnamefont {Nicolas}\ \bibnamefont
  {Regnault}}, \bibinfo {author} {\bibfnamefont {Yuanfeng}\ \bibnamefont {Xu}},
  \bibinfo {author} {\bibfnamefont {Ming-Rui}\ \bibnamefont {Li}}, \bibinfo
  {author} {\bibfnamefont {Da-Shuai}\ \bibnamefont {Ma}}, \bibinfo {author}
  {\bibfnamefont {Milena}\ \bibnamefont {Jovanovic}}, \bibinfo {author}
  {\bibfnamefont {Ali}\ \bibnamefont {Yazdani}}, \bibinfo {author}
  {\bibfnamefont {Stuart S.~P.}\ \bibnamefont {Parkin}}, \bibinfo {author}
  {\bibfnamefont {Claudia}\ \bibnamefont {Felser}}, \bibinfo {author}
  {\bibfnamefont {Leslie~M.}\ \bibnamefont {Schoop}}, \bibinfo {author}
  {\bibfnamefont {N.~Phuan}\ \bibnamefont {Ong}}, \bibinfo {author}
  {\bibfnamefont {Robert~J.}\ \bibnamefont {Cava}}, \bibinfo {author}
  {\bibfnamefont {Luis}\ \bibnamefont {Elcoro}}, \bibinfo {author}
  {\bibfnamefont {Zhi-Da}\ \bibnamefont {Song}}, \ and\ \bibinfo {author}
  {\bibfnamefont {B.~Andrei}\ \bibnamefont {Bernevig}},\ }\bibfield  {title}
  {\enquote {\bibinfo {title} {{Catalogue of flat-band stoichiometric
  materials}},}\ }\href {\doibase 10.1038/s41586-022-04519-1} {\bibfield
  {journal} {\bibinfo  {journal} {Nature}\ }\textbf {\bibinfo {volume} {603}},\
  \bibinfo {pages} {824--828} (\bibinfo {year} {2022})}\BibitemShut {NoStop}%
\bibitem [{\citenamefont {Ma}\ \emph {et~al.}(2020)\citenamefont {Ma},
  \citenamefont {Xu}, \citenamefont {Chiu}, \citenamefont {Regnault},
  \citenamefont {Houck}, \citenamefont {Song},\ and\ \citenamefont
  {Bernevig}}]{Ma2020}%
  \BibitemOpen
  \bibfield  {author} {\bibinfo {author} {\bibfnamefont {Da-Shuai}\
  \bibnamefont {Ma}}, \bibinfo {author} {\bibfnamefont {Yuanfeng}\ \bibnamefont
  {Xu}}, \bibinfo {author} {\bibfnamefont {Christie~S.}\ \bibnamefont {Chiu}},
  \bibinfo {author} {\bibfnamefont {Nicolas}\ \bibnamefont {Regnault}},
  \bibinfo {author} {\bibfnamefont {Andrew~A.}\ \bibnamefont {Houck}}, \bibinfo
  {author} {\bibfnamefont {Zhida}\ \bibnamefont {Song}}, \ and\ \bibinfo
  {author} {\bibfnamefont {B.~Andrei}\ \bibnamefont {Bernevig}},\ }\bibfield
  {title} {\enquote {\bibinfo {title} {Spin-orbit-induced topological flat
  bands in line and split graphs of bipartite lattices},}\ }\href {\doibase
  10.1103/PhysRevLett.125.266403} {\bibfield  {journal} {\bibinfo  {journal}
  {Physical Review Letters}\ }\textbf {\bibinfo {volume} {125}},\ \bibinfo
  {pages} {266403} (\bibinfo {year} {2020})}\BibitemShut {NoStop}%
\bibitem [{\citenamefont {Yin}\ \emph {et~al.}(2022)\citenamefont {Yin},
  \citenamefont {Lian},\ and\ \citenamefont {Hasan}}]{Yin2022}%
  \BibitemOpen
  \bibfield  {author} {\bibinfo {author} {\bibfnamefont {Jia-Xin}\ \bibnamefont
  {Yin}}, \bibinfo {author} {\bibfnamefont {Biao}\ \bibnamefont {Lian}}, \ and\
  \bibinfo {author} {\bibfnamefont {M.~Zahid}\ \bibnamefont {Hasan}},\
  }\bibfield  {title} {\enquote {\bibinfo {title} {Topological kagome magnets
  and superconductors},}\ }\href {\doibase 10.1038/s41586-022-05516-0}
  {\bibfield  {journal} {\bibinfo  {journal} {Nature}\ }\textbf {\bibinfo
  {volume} {612}},\ \bibinfo {pages} {647--657} (\bibinfo {year}
  {2022})}\BibitemShut {NoStop}%
\bibitem [{\citenamefont {Tang}\ \emph {et~al.}(2011)\citenamefont {Tang},
  \citenamefont {Mei},\ and\ \citenamefont {Wen}}]{Wen2011}%
  \BibitemOpen
  \bibfield  {author} {\bibinfo {author} {\bibfnamefont {Evelyn}\ \bibnamefont
  {Tang}}, \bibinfo {author} {\bibfnamefont {Jia-Wei}\ \bibnamefont {Mei}}, \
  and\ \bibinfo {author} {\bibfnamefont {Xiao-Gang}\ \bibnamefont {Wen}},\
  }\bibfield  {title} {\enquote {\bibinfo {title} {High-temperature fractional
  quantum hall states},}\ }\href {\doibase 10.1103/PhysRevLett.106.236802}
  {\bibfield  {journal} {\bibinfo  {journal} {Phys. Rev. Lett.}\ }\textbf
  {\bibinfo {volume} {106}},\ \bibinfo {pages} {236802} (\bibinfo {year}
  {2011})}\BibitemShut {NoStop}%
\bibitem [{\citenamefont {Sun}\ \emph {et~al.}(2011)\citenamefont {Sun},
  \citenamefont {Gu}, \citenamefont {Katsura},\ and\ \citenamefont
  {Das~Sarma}}]{SunKai2011}%
  \BibitemOpen
  \bibfield  {author} {\bibinfo {author} {\bibfnamefont {Kai}\ \bibnamefont
  {Sun}}, \bibinfo {author} {\bibfnamefont {Zhengcheng}\ \bibnamefont {Gu}},
  \bibinfo {author} {\bibfnamefont {Hosho}\ \bibnamefont {Katsura}}, \ and\
  \bibinfo {author} {\bibfnamefont {S.}~\bibnamefont {Das~Sarma}},\ }\bibfield
  {title} {\enquote {\bibinfo {title} {Nearly flatbands with nontrivial
  topology},}\ }\href {\doibase 10.1103/PhysRevLett.106.236803} {\bibfield
  {journal} {\bibinfo  {journal} {Phys. Rev. Lett.}\ }\textbf {\bibinfo
  {volume} {106}},\ \bibinfo {pages} {236803} (\bibinfo {year}
  {2011})}\BibitemShut {NoStop}%
\bibitem [{\citenamefont {Cao}\ \emph {et~al.}(2018)\citenamefont {Cao},
  \citenamefont {Fatemi}, \citenamefont {Fang}, \citenamefont {Watanabe},
  \citenamefont {Taniguchi}, \citenamefont {Kaxiras},\ and\ \citenamefont
  {Jarillo-Herrero}}]{Cao2018}%
  \BibitemOpen
  \bibfield  {author} {\bibinfo {author} {\bibfnamefont {Yuan}\ \bibnamefont
  {Cao}}, \bibinfo {author} {\bibfnamefont {Valla}\ \bibnamefont {Fatemi}},
  \bibinfo {author} {\bibfnamefont {Shiang}\ \bibnamefont {Fang}}, \bibinfo
  {author} {\bibfnamefont {Kenji}\ \bibnamefont {Watanabe}}, \bibinfo {author}
  {\bibfnamefont {Takashi}\ \bibnamefont {Taniguchi}}, \bibinfo {author}
  {\bibfnamefont {Efthimios}\ \bibnamefont {Kaxiras}}, \ and\ \bibinfo {author}
  {\bibfnamefont {Pablo}\ \bibnamefont {Jarillo-Herrero}},\ }\bibfield  {title}
  {\enquote {\bibinfo {title} {Unconventional superconductivity in magic-angle
  graphene superlattices},}\ }\href {\doibase 10.1038/nature26160} {\bibfield
  {journal} {\bibinfo  {journal} {Nature}\ }\textbf {\bibinfo {volume} {556}},\
  \bibinfo {pages} {43--50} (\bibinfo {year} {2018})}\BibitemShut {NoStop}%
\bibitem [{\citenamefont {Bistritzer}\ and\ \citenamefont
  {MacDonald}(2011)}]{Bistritzer2011}%
  \BibitemOpen
  \bibfield  {author} {\bibinfo {author} {\bibfnamefont {Rafi}\ \bibnamefont
  {Bistritzer}}\ and\ \bibinfo {author} {\bibfnamefont {Allan~H.}\ \bibnamefont
  {MacDonald}},\ }\bibfield  {title} {\enquote {\bibinfo {title} {Moiré bands
  in twisted double-layer graphene},}\ }\href {\doibase
  10.1073/pnas.1108174108} {\bibfield  {journal} {\bibinfo  {journal}
  {Proceedings of the National Academy of Sciences}\ }\textbf {\bibinfo
  {volume} {108}},\ \bibinfo {pages} {12233--12237} (\bibinfo {year}
  {2011})}\BibitemShut {NoStop}%
\bibitem [{\citenamefont {Chen}\ \emph {et~al.}(2021)\citenamefont {Chen},
  \citenamefont {Yang}, \citenamefont {Hu}, \citenamefont {Zhao}, \citenamefont
  {Yuan}, \citenamefont {Xing}, \citenamefont {Qian}, \citenamefont {Huang},
  \citenamefont {Li}, \citenamefont {Ye}, \citenamefont {Ma}, \citenamefont
  {Ni}, \citenamefont {Zhang}, \citenamefont {Yin}, \citenamefont {Gong},
  \citenamefont {Tu}, \citenamefont {Lei}, \citenamefont {Tan}, \citenamefont
  {Zhou}, \citenamefont {Shen}, \citenamefont {Dong}, \citenamefont {Yan},
  \citenamefont {Wang},\ and\ \citenamefont {Gao}}]{Chen2021}%
  \BibitemOpen
  \bibfield  {author} {\bibinfo {author} {\bibfnamefont {Hui}\ \bibnamefont
  {Chen}}, \bibinfo {author} {\bibfnamefont {Haitao}\ \bibnamefont {Yang}},
  \bibinfo {author} {\bibfnamefont {Bin}\ \bibnamefont {Hu}}, \bibinfo {author}
  {\bibfnamefont {Zhen}\ \bibnamefont {Zhao}}, \bibinfo {author} {\bibfnamefont
  {Jie}\ \bibnamefont {Yuan}}, \bibinfo {author} {\bibfnamefont {Yuqing}\
  \bibnamefont {Xing}}, \bibinfo {author} {\bibfnamefont {Guojian}\
  \bibnamefont {Qian}}, \bibinfo {author} {\bibfnamefont {Zihao}\ \bibnamefont
  {Huang}}, \bibinfo {author} {\bibfnamefont {Geng}\ \bibnamefont {Li}},
  \bibinfo {author} {\bibfnamefont {Yuhan}\ \bibnamefont {Ye}}, \bibinfo
  {author} {\bibfnamefont {Sheng}\ \bibnamefont {Ma}}, \bibinfo {author}
  {\bibfnamefont {Shunli}\ \bibnamefont {Ni}}, \bibinfo {author} {\bibfnamefont
  {Hua}\ \bibnamefont {Zhang}}, \bibinfo {author} {\bibfnamefont {Qiangwei}\
  \bibnamefont {Yin}}, \bibinfo {author} {\bibfnamefont {Chunsheng}\
  \bibnamefont {Gong}}, \bibinfo {author} {\bibfnamefont {Zhijun}\ \bibnamefont
  {Tu}}, \bibinfo {author} {\bibfnamefont {Hechang}\ \bibnamefont {Lei}},
  \bibinfo {author} {\bibfnamefont {Hengxin}\ \bibnamefont {Tan}}, \bibinfo
  {author} {\bibfnamefont {Sen}\ \bibnamefont {Zhou}}, \bibinfo {author}
  {\bibfnamefont {Chengmin}\ \bibnamefont {Shen}}, \bibinfo {author}
  {\bibfnamefont {Xiaoli}\ \bibnamefont {Dong}}, \bibinfo {author}
  {\bibfnamefont {Binghai}\ \bibnamefont {Yan}}, \bibinfo {author}
  {\bibfnamefont {Ziqiang}\ \bibnamefont {Wang}}, \ and\ \bibinfo {author}
  {\bibfnamefont {Hong-Jun}\ \bibnamefont {Gao}},\ }\bibfield  {title}
  {\enquote {\bibinfo {title} {Roton pair density wave in a strong-coupling
  kagome superconductor},}\ }\href {\doibase 10.1038/s41586-021-03983-5}
  {\bibfield  {journal} {\bibinfo  {journal} {Nature}\ }\textbf {\bibinfo
  {volume} {599}},\ \bibinfo {pages} {222--228} (\bibinfo {year}
  {2021})}\BibitemShut {NoStop}%
\bibitem [{\citenamefont {Kang}\ \emph {et~al.}(2022)\citenamefont {Kang},
  \citenamefont {Fang}, \citenamefont {Yoo}, \citenamefont {Ortiz},
  \citenamefont {Oey}, \citenamefont {Choi}, \citenamefont {Ryu}, \citenamefont
  {Kim}, \citenamefont {Jozwiak}, \citenamefont {Bostwick}, \citenamefont
  {Rotenberg}, \citenamefont {Kaxiras}, \citenamefont {Checkelsky},
  \citenamefont {Wilson}, \citenamefont {Park},\ and\ \citenamefont
  {Comin}}]{Kang2022}%
  \BibitemOpen
  \bibfield  {author} {\bibinfo {author} {\bibfnamefont {Mingu}\ \bibnamefont
  {Kang}}, \bibinfo {author} {\bibfnamefont {Shiang}\ \bibnamefont {Fang}},
  \bibinfo {author} {\bibfnamefont {Jonggyu}\ \bibnamefont {Yoo}}, \bibinfo
  {author} {\bibfnamefont {Brenden~R.}\ \bibnamefont {Ortiz}}, \bibinfo
  {author} {\bibfnamefont {Yuzki~M.}\ \bibnamefont {Oey}}, \bibinfo {author}
  {\bibfnamefont {Jonghyeok}\ \bibnamefont {Choi}}, \bibinfo {author}
  {\bibfnamefont {Sae~Hee}\ \bibnamefont {Ryu}}, \bibinfo {author}
  {\bibfnamefont {Jimin}\ \bibnamefont {Kim}}, \bibinfo {author} {\bibfnamefont
  {Chris}\ \bibnamefont {Jozwiak}}, \bibinfo {author} {\bibfnamefont {Aaron}\
  \bibnamefont {Bostwick}}, \bibinfo {author} {\bibfnamefont {Eli}\
  \bibnamefont {Rotenberg}}, \bibinfo {author} {\bibfnamefont {Efthimios}\
  \bibnamefont {Kaxiras}}, \bibinfo {author} {\bibfnamefont {Joseph~G.}\
  \bibnamefont {Checkelsky}}, \bibinfo {author} {\bibfnamefont {Stephen~D.}\
  \bibnamefont {Wilson}}, \bibinfo {author} {\bibfnamefont {Jae-Hoon}\
  \bibnamefont {Park}}, \ and\ \bibinfo {author} {\bibfnamefont {Riccardo}\
  \bibnamefont {Comin}},\ }\bibfield  {title} {\enquote {\bibinfo {title}
  {Charge order landscape and competition with superconductivity in kagome
  metals},}\ }\href {\doibase 10.1038/s41563-022-01375-2} {\bibfield  {journal}
  {\bibinfo  {journal} {Nature Materials}\ }\textbf {\bibinfo {volume} {22}},\
  \bibinfo {pages} {186--193} (\bibinfo {year} {2022})}\BibitemShut {NoStop}%
\bibitem [{\citenamefont {Jiang}\ \emph {et~al.}(2023)\citenamefont {Jiang},
  \citenamefont {Liu}, \citenamefont {Ma}, \citenamefont {Xia}, \citenamefont
  {Liu}, \citenamefont {Liu}, \citenamefont {Cho}, \citenamefont {Yang},
  \citenamefont {Ding}, \citenamefont {Liu}, \citenamefont {Huang},
  \citenamefont {Qiao}, \citenamefont {Shen}, \citenamefont {Jing},
  \citenamefont {Liu}, \citenamefont {Liu}, \citenamefont {Guo},\ and\
  \citenamefont {Shen}}]{Jiang2023}%
  \BibitemOpen
  \bibfield  {author} {\bibinfo {author} {\bibfnamefont {Zhicheng}\
  \bibnamefont {Jiang}}, \bibinfo {author} {\bibfnamefont {Zhengtai}\
  \bibnamefont {Liu}}, \bibinfo {author} {\bibfnamefont {Haiyang}\ \bibnamefont
  {Ma}}, \bibinfo {author} {\bibfnamefont {Wei}\ \bibnamefont {Xia}}, \bibinfo
  {author} {\bibfnamefont {Zhonghao}\ \bibnamefont {Liu}}, \bibinfo {author}
  {\bibfnamefont {Jishan}\ \bibnamefont {Liu}}, \bibinfo {author}
  {\bibfnamefont {Soohyun}\ \bibnamefont {Cho}}, \bibinfo {author}
  {\bibfnamefont {Yichen}\ \bibnamefont {Yang}}, \bibinfo {author}
  {\bibfnamefont {Jianyang}\ \bibnamefont {Ding}}, \bibinfo {author}
  {\bibfnamefont {Jiayu}\ \bibnamefont {Liu}}, \bibinfo {author} {\bibfnamefont
  {Zhe}\ \bibnamefont {Huang}}, \bibinfo {author} {\bibfnamefont {Yuxi}\
  \bibnamefont {Qiao}}, \bibinfo {author} {\bibfnamefont {Jiajia}\ \bibnamefont
  {Shen}}, \bibinfo {author} {\bibfnamefont {Wenchuan}\ \bibnamefont {Jing}},
  \bibinfo {author} {\bibfnamefont {Xiangqi}\ \bibnamefont {Liu}}, \bibinfo
  {author} {\bibfnamefont {Jianpeng}\ \bibnamefont {Liu}}, \bibinfo {author}
  {\bibfnamefont {Yanfeng}\ \bibnamefont {Guo}}, \ and\ \bibinfo {author}
  {\bibfnamefont {Dawei}\ \bibnamefont {Shen}},\ }\bibfield  {title} {\enquote
  {\bibinfo {title} {Flat bands, non-trivial band topology and rotation
  symmetry breaking in layered kagome-lattice rbti3bi5},}\ }\href {\doibase
  10.1038/s41467-023-40515-3} {\bibfield  {journal} {\bibinfo  {journal}
  {Nature Communications}\ }\textbf {\bibinfo {volume} {14}},\ \bibinfo {pages}
  {4892} (\bibinfo {year} {2023})}\BibitemShut {NoStop}%
\bibitem [{\citenamefont {Singh}(2023)}]{Singh2023}%
  \BibitemOpen
  \bibfield  {author} {\bibinfo {author} {\bibfnamefont {Bahadur}\ \bibnamefont
  {Singh}},\ }\bibfield  {title} {\enquote {\bibinfo {title} {Rotation
  rearranges electrons},}\ }\href {\doibase 10.1038/s41567-023-02237-7}
  {\bibfield  {journal} {\bibinfo  {journal} {Nature Physics}\ } (\bibinfo
  {year} {2023}),\ 10.1038/s41567-023-02237-7}\BibitemShut {NoStop}%
\bibitem [{\citenamefont {Wang}\ \emph {et~al.}(2023)\citenamefont {Wang},
  \citenamefont {Wu}, \citenamefont {McCandless}, \citenamefont {Chan},\ and\
  \citenamefont {Ali}}]{Wang2023}%
  \BibitemOpen
  \bibfield  {author} {\bibinfo {author} {\bibfnamefont {Yaojia}\ \bibnamefont
  {Wang}}, \bibinfo {author} {\bibfnamefont {Heng}\ \bibnamefont {Wu}},
  \bibinfo {author} {\bibfnamefont {Gregory~T.}\ \bibnamefont {McCandless}},
  \bibinfo {author} {\bibfnamefont {Julia~Y.}\ \bibnamefont {Chan}}, \ and\
  \bibinfo {author} {\bibfnamefont {Mazhar~N.}\ \bibnamefont {Ali}},\
  }\bibfield  {title} {\enquote {\bibinfo {title} {Quantum states and
  intertwining phases in kagome materials},}\ }\href {\doibase
  10.1038/s42254-023-00635-7} {\bibfield  {journal} {\bibinfo  {journal}
  {Nature Reviews Physics}\ } (\bibinfo {year} {2023}),\
  10.1038/s42254-023-00635-7}\BibitemShut {NoStop}%
\bibitem [{\citenamefont {Kang}\ \emph {et~al.}(2020)\citenamefont {Kang},
  \citenamefont {Fang}, \citenamefont {Ye}, \citenamefont {Po}, \citenamefont
  {Denlinger}, \citenamefont {Jozwiak}, \citenamefont {Bostwick}, \citenamefont
  {Rotenberg}, \citenamefont {Kaxiras}, \citenamefont {Checkelsky},\ and\
  \citenamefont {Comin}}]{Kang2020}%
  \BibitemOpen
  \bibfield  {author} {\bibinfo {author} {\bibfnamefont {Mingu}\ \bibnamefont
  {Kang}}, \bibinfo {author} {\bibfnamefont {Shiang}\ \bibnamefont {Fang}},
  \bibinfo {author} {\bibfnamefont {Linda}\ \bibnamefont {Ye}}, \bibinfo
  {author} {\bibfnamefont {Hoi~Chun}\ \bibnamefont {Po}}, \bibinfo {author}
  {\bibfnamefont {Jonathan}\ \bibnamefont {Denlinger}}, \bibinfo {author}
  {\bibfnamefont {Chris}\ \bibnamefont {Jozwiak}}, \bibinfo {author}
  {\bibfnamefont {Aaron}\ \bibnamefont {Bostwick}}, \bibinfo {author}
  {\bibfnamefont {Eli}\ \bibnamefont {Rotenberg}}, \bibinfo {author}
  {\bibfnamefont {Efthimios}\ \bibnamefont {Kaxiras}}, \bibinfo {author}
  {\bibfnamefont {Joseph~G.}\ \bibnamefont {Checkelsky}}, \ and\ \bibinfo
  {author} {\bibfnamefont {Riccardo}\ \bibnamefont {Comin}},\ }\bibfield
  {title} {\enquote {\bibinfo {title} {{Topological flat bands in frustrated
  kagome lattice CoSn}},}\ }\href {\doibase 10.1038/s41467-020-17465-1}
  {\bibfield  {journal} {\bibinfo  {journal} {Nature Communications}\ }\textbf
  {\bibinfo {volume} {11}},\ \bibinfo {pages} {1--9} (\bibinfo {year}
  {2020})}\BibitemShut {NoStop}%
\bibitem [{\citenamefont {Markiewicz}\ \emph {et~al.}(2023)\citenamefont
  {Markiewicz}, \citenamefont {Singh}, \citenamefont {Lane},\ and\
  \citenamefont {Bansil}}]{Markiewicz2023}%
  \BibitemOpen
  \bibfield  {author} {\bibinfo {author} {\bibfnamefont {Robert~S.}\
  \bibnamefont {Markiewicz}}, \bibinfo {author} {\bibfnamefont {Bahadur}\
  \bibnamefont {Singh}}, \bibinfo {author} {\bibfnamefont {Christopher}\
  \bibnamefont {Lane}}, \ and\ \bibinfo {author} {\bibfnamefont {Arun}\
  \bibnamefont {Bansil}},\ }\bibfield  {title} {\enquote {\bibinfo {title}
  {Investigating the cuprates as a platform for high-order van hove
  singularities and flat-band physics},}\ }\href {\doibase
  10.1038/s42005-023-01373-z} {\bibfield  {journal} {\bibinfo  {journal}
  {Communications Physics}\ }\textbf {\bibinfo {volume} {6}},\ \bibinfo {pages}
  {292} (\bibinfo {year} {2023})}\BibitemShut {NoStop}%
\bibitem [{\citenamefont {Li}\ \emph {et~al.}(2019)\citenamefont {Li},
  \citenamefont {Lee}, \citenamefont {Wang}, \citenamefont {Osada},
  \citenamefont {Crossley}, \citenamefont {Lee}, \citenamefont {Cui},
  \citenamefont {Hikita},\ and\ \citenamefont {Hwang}}]{Li2019a}%
  \BibitemOpen
  \bibfield  {author} {\bibinfo {author} {\bibfnamefont {Danfeng}\ \bibnamefont
  {Li}}, \bibinfo {author} {\bibfnamefont {Kyuho}\ \bibnamefont {Lee}},
  \bibinfo {author} {\bibfnamefont {Bai~Yang}\ \bibnamefont {Wang}}, \bibinfo
  {author} {\bibfnamefont {Motoki}\ \bibnamefont {Osada}}, \bibinfo {author}
  {\bibfnamefont {Samuel}\ \bibnamefont {Crossley}}, \bibinfo {author}
  {\bibfnamefont {Hye~Ryoung}\ \bibnamefont {Lee}}, \bibinfo {author}
  {\bibfnamefont {Yi}~\bibnamefont {Cui}}, \bibinfo {author} {\bibfnamefont
  {Yasuyuki}\ \bibnamefont {Hikita}}, \ and\ \bibinfo {author} {\bibfnamefont
  {Harold~Y.}\ \bibnamefont {Hwang}},\ }\bibfield  {title} {\enquote {\bibinfo
  {title} {{Superconductivity in an infinite-layer nickelate}},}\ }\href
  {\doibase 10.1038/s41586-019-1496-5} {\bibfield  {journal} {\bibinfo
  {journal} {Nature}\ }\textbf {\bibinfo {volume} {572}},\ \bibinfo {pages}
  {624--627} (\bibinfo {year} {2019})}\BibitemShut {NoStop}%
\bibitem [{\citenamefont {Osada}\ \emph {et~al.}(2020)\citenamefont {Osada},
  \citenamefont {Wang}, \citenamefont {Goodge}, \citenamefont {Lee},
  \citenamefont {Yoon}, \citenamefont {Sakuma}, \citenamefont {Li},
  \citenamefont {Miura}, \citenamefont {Kourkoutis},\ and\ \citenamefont
  {Hwang}}]{Osada2020a}%
  \BibitemOpen
  \bibfield  {author} {\bibinfo {author} {\bibfnamefont {Motoki}\ \bibnamefont
  {Osada}}, \bibinfo {author} {\bibfnamefont {Bai~Yang}\ \bibnamefont {Wang}},
  \bibinfo {author} {\bibfnamefont {Berit~H.}\ \bibnamefont {Goodge}}, \bibinfo
  {author} {\bibfnamefont {Kyuho}\ \bibnamefont {Lee}}, \bibinfo {author}
  {\bibfnamefont {Hyeok}\ \bibnamefont {Yoon}}, \bibinfo {author}
  {\bibfnamefont {Keita}\ \bibnamefont {Sakuma}}, \bibinfo {author}
  {\bibfnamefont {Danfeng}\ \bibnamefont {Li}}, \bibinfo {author}
  {\bibfnamefont {Masashi}\ \bibnamefont {Miura}}, \bibinfo {author}
  {\bibfnamefont {Lena~F.}\ \bibnamefont {Kourkoutis}}, \ and\ \bibinfo
  {author} {\bibfnamefont {Harold~Y.}\ \bibnamefont {Hwang}},\ }\bibfield
  {title} {\enquote {\bibinfo {title} {{A Superconducting Praseodymium
  Nickelate with Infinite Layer Structure}},}\ }\href {\doibase
  10.1021/acs.nanolett.0c01392} {\bibfield  {journal} {\bibinfo  {journal}
  {Nano Letters}\ }\textbf {\bibinfo {volume} {20}},\ \bibinfo {pages}
  {5735--5740} (\bibinfo {year} {2020})}\BibitemShut {NoStop}%
\bibitem [{\citenamefont {Zeng}\ \emph {et~al.}(2022)\citenamefont {Zeng},
  \citenamefont {Li}, \citenamefont {Chow}, \citenamefont {Cao}, \citenamefont
  {Zhang}, \citenamefont {Tang}, \citenamefont {Yin}, \citenamefont {Lim},
  \citenamefont {Hu}, \citenamefont {Yang},\ and\ \citenamefont
  {Ariando}}]{ZhangS2021}%
  \BibitemOpen
  \bibfield  {author} {\bibinfo {author} {\bibfnamefont {Shengwei}\
  \bibnamefont {Zeng}}, \bibinfo {author} {\bibfnamefont {Changjian}\
  \bibnamefont {Li}}, \bibinfo {author} {\bibfnamefont {Lin~Er}\ \bibnamefont
  {Chow}}, \bibinfo {author} {\bibfnamefont {Yu}~\bibnamefont {Cao}}, \bibinfo
  {author} {\bibfnamefont {Zhaoting}\ \bibnamefont {Zhang}}, \bibinfo {author}
  {\bibfnamefont {Chi~Sin}\ \bibnamefont {Tang}}, \bibinfo {author}
  {\bibfnamefont {Xinmao}\ \bibnamefont {Yin}}, \bibinfo {author}
  {\bibfnamefont {Zhi~Shiuh}\ \bibnamefont {Lim}}, \bibinfo {author}
  {\bibfnamefont {Junxiong}\ \bibnamefont {Hu}}, \bibinfo {author}
  {\bibfnamefont {Ping}\ \bibnamefont {Yang}}, \ and\ \bibinfo {author}
  {\bibfnamefont {Ariando}\ \bibnamefont {Ariando}},\ }\bibfield  {title}
  {\enquote {\bibinfo {title} {Superconductivity in infinite-layer nickelate
  la$_{1-x}$ca$_x$nio$_2$ thin films},}\ }\href {\doibase
  10.1126/sciadv.abl9927} {\bibfield  {journal} {\bibinfo  {journal} {Science
  Advances}\ }\textbf {\bibinfo {volume} {8}},\ \bibinfo {pages} {9927}
  (\bibinfo {year} {2022})}\BibitemShut {NoStop}%
\bibitem [{\citenamefont {Osada}\ \emph
  {et~al.}(2021{\natexlab{a}})\citenamefont {Osada}, \citenamefont {Wang},
  \citenamefont {Goodge}, \citenamefont {Harvey}, \citenamefont {Lee},
  \citenamefont {Li}, \citenamefont {Kourkoutis},\ and\ \citenamefont
  {Hwang}}]{Osada2021}%
  \BibitemOpen
  \bibfield  {author} {\bibinfo {author} {\bibfnamefont {Motoki}\ \bibnamefont
  {Osada}}, \bibinfo {author} {\bibfnamefont {Bai~Yang}\ \bibnamefont {Wang}},
  \bibinfo {author} {\bibfnamefont {Berit~H.}\ \bibnamefont {Goodge}}, \bibinfo
  {author} {\bibfnamefont {Shannon~P.}\ \bibnamefont {Harvey}}, \bibinfo
  {author} {\bibfnamefont {Kyuho}\ \bibnamefont {Lee}}, \bibinfo {author}
  {\bibfnamefont {Danfeng}\ \bibnamefont {Li}}, \bibinfo {author}
  {\bibfnamefont {Lena~F.}\ \bibnamefont {Kourkoutis}}, \ and\ \bibinfo
  {author} {\bibfnamefont {Harold~Y.}\ \bibnamefont {Hwang}},\ }\bibfield
  {title} {\enquote {\bibinfo {title} {{Nickelate Superconductivity without
  Rare‐Earth Magnetism: (La,Sr)NiO$_2$}},}\ }\href {\doibase
  10.1002/adma.202104083} {\bibfield  {journal} {\bibinfo  {journal} {Advanced
  Materials}\ }\textbf {\bibinfo {volume} {33}} (\bibinfo {year}
  {2021}{\natexlab{a}}),\ 10.1002/adma.202104083}\BibitemShut {NoStop}%
\bibitem [{\citenamefont {Sawatzky}(2019)}]{Sawatzky2019}%
  \BibitemOpen
  \bibfield  {author} {\bibinfo {author} {\bibfnamefont {George~A.}\
  \bibnamefont {Sawatzky}},\ }\bibfield  {title} {\enquote {\bibinfo {title}
  {{Superconductivity seen in a non-magnetic nickel oxide}},}\ }\href {\doibase
  10.1038/d41586-019-02518-3} {\bibfield  {journal} {\bibinfo  {journal}
  {Nature}\ }\textbf {\bibinfo {volume} {572}},\ \bibinfo {pages} {592--593}
  (\bibinfo {year} {2019})}\BibitemShut {NoStop}%
\bibitem [{\citenamefont {Norman}(2020)}]{Norman2020}%
  \BibitemOpen
  \bibfield  {author} {\bibinfo {author} {\bibfnamefont {Michael~R.}\
  \bibnamefont {Norman}},\ }\bibfield  {title} {\enquote {\bibinfo {title}
  {{Entering the Nickel Age of Superconductivity}},}\ }\href {\doibase
  10.1103/Physics.13.85} {\bibfield  {journal} {\bibinfo  {journal} {Physics}\
  }\textbf {\bibinfo {volume} {13}},\ \bibinfo {pages} {85} (\bibinfo {year}
  {2020})}\BibitemShut {NoStop}%
\bibitem [{\citenamefont {Pickett}(2021)}]{Pickett2021}%
  \BibitemOpen
  \bibfield  {author} {\bibinfo {author} {\bibfnamefont {Warren~E.}\
  \bibnamefont {Pickett}},\ }\bibfield  {title} {\enquote {\bibinfo {title}
  {{The dawn of the nickel age of superconductivity}},}\ }\href {\doibase
  10.1038/s42254-020-00257-3} {\bibfield  {journal} {\bibinfo  {journal}
  {Nature Reviews Physics}\ }\textbf {\bibinfo {volume} {3}},\ \bibinfo {pages}
  {7--8} (\bibinfo {year} {2021})}\BibitemShut {NoStop}%
\bibitem [{\citenamefont {Hepting}\ \emph {et~al.}(2020)\citenamefont
  {Hepting}, \citenamefont {Li}, \citenamefont {Jia}, \citenamefont {Lu},
  \citenamefont {Paris}, \citenamefont {Tseng}, \citenamefont {Feng},
  \citenamefont {Osada}, \citenamefont {Been}, \citenamefont {Hikita},
  \citenamefont {Chuang}, \citenamefont {Hussain}, \citenamefont {Zhou},
  \citenamefont {Nag}, \citenamefont {Garcia-Fernandez}, \citenamefont {Rossi},
  \citenamefont {Huang}, \citenamefont {Huang}, \citenamefont {Shen},
  \citenamefont {Schmitt}, \citenamefont {Hwang}, \citenamefont {Moritz},
  \citenamefont {Zaanen}, \citenamefont {Devereaux},\ and\ \citenamefont
  {Lee}}]{Hepting2020}%
  \BibitemOpen
  \bibfield  {author} {\bibinfo {author} {\bibfnamefont {M.}~\bibnamefont
  {Hepting}}, \bibinfo {author} {\bibfnamefont {D.}~\bibnamefont {Li}},
  \bibinfo {author} {\bibfnamefont {C.~J.}\ \bibnamefont {Jia}}, \bibinfo
  {author} {\bibfnamefont {H.}~\bibnamefont {Lu}}, \bibinfo {author}
  {\bibfnamefont {E.}~\bibnamefont {Paris}}, \bibinfo {author} {\bibfnamefont
  {Y.}~\bibnamefont {Tseng}}, \bibinfo {author} {\bibfnamefont
  {X.}~\bibnamefont {Feng}}, \bibinfo {author} {\bibfnamefont {M.}~\bibnamefont
  {Osada}}, \bibinfo {author} {\bibfnamefont {E.}~\bibnamefont {Been}},
  \bibinfo {author} {\bibfnamefont {Y.}~\bibnamefont {Hikita}}, \bibinfo
  {author} {\bibfnamefont {Y.-D.}\ \bibnamefont {Chuang}}, \bibinfo {author}
  {\bibfnamefont {Z.}~\bibnamefont {Hussain}}, \bibinfo {author} {\bibfnamefont
  {K.~J.}\ \bibnamefont {Zhou}}, \bibinfo {author} {\bibfnamefont
  {A.}~\bibnamefont {Nag}}, \bibinfo {author} {\bibfnamefont {M.}~\bibnamefont
  {Garcia-Fernandez}}, \bibinfo {author} {\bibfnamefont {M.}~\bibnamefont
  {Rossi}}, \bibinfo {author} {\bibfnamefont {H.~Y.}\ \bibnamefont {Huang}},
  \bibinfo {author} {\bibfnamefont {D.~J.}\ \bibnamefont {Huang}}, \bibinfo
  {author} {\bibfnamefont {Z.~X.}\ \bibnamefont {Shen}}, \bibinfo {author}
  {\bibfnamefont {T.}~\bibnamefont {Schmitt}}, \bibinfo {author} {\bibfnamefont
  {H.~Y.}\ \bibnamefont {Hwang}}, \bibinfo {author} {\bibfnamefont
  {B.}~\bibnamefont {Moritz}}, \bibinfo {author} {\bibfnamefont
  {J.}~\bibnamefont {Zaanen}}, \bibinfo {author} {\bibfnamefont {T.~P.}\
  \bibnamefont {Devereaux}}, \ and\ \bibinfo {author} {\bibfnamefont {W.~S.}\
  \bibnamefont {Lee}},\ }\bibfield  {title} {\enquote {\bibinfo {title}
  {{Electronic structure of the parent compound of superconducting
  infinite-layer nickelates}},}\ }\href {\doibase 10.1038/s41563-019-0585-z}
  {\bibfield  {journal} {\bibinfo  {journal} {Nature Materials}\ }\textbf
  {\bibinfo {volume} {19}},\ \bibinfo {pages} {381--385} (\bibinfo {year}
  {2020})}\BibitemShut {NoStop}%
\bibitem [{\citenamefont {Liu}\ \emph {et~al.}(2020)\citenamefont {Liu},
  \citenamefont {Ren}, \citenamefont {Zhu}, \citenamefont {Wang},\ and\
  \citenamefont {Yang}}]{Liu2020}%
  \BibitemOpen
  \bibfield  {author} {\bibinfo {author} {\bibfnamefont {Zhao}\ \bibnamefont
  {Liu}}, \bibinfo {author} {\bibfnamefont {Zhi}\ \bibnamefont {Ren}}, \bibinfo
  {author} {\bibfnamefont {Wei}\ \bibnamefont {Zhu}}, \bibinfo {author}
  {\bibfnamefont {Zhengfei}\ \bibnamefont {Wang}}, \ and\ \bibinfo {author}
  {\bibfnamefont {Jinlong}\ \bibnamefont {Yang}},\ }\bibfield  {title}
  {\enquote {\bibinfo {title} {{Electronic and magnetic structure of
  infinite-layer NdNiO2: trace of antiferromagnetic metal}},}\ }\href {\doibase
  10.1038/s41535-020-0229-1} {\bibfield  {journal} {\bibinfo  {journal} {npj
  Quantum Materials}\ }\textbf {\bibinfo {volume} {5}},\ \bibinfo {pages} {31}
  (\bibinfo {year} {2020})}\BibitemShut {NoStop}%
\bibitem [{\citenamefont {Choi}\ \emph
  {et~al.}(2020{\natexlab{a}})\citenamefont {Choi}, \citenamefont {Lee},\ and\
  \citenamefont {Pickett}}]{Choi2020}%
  \BibitemOpen
  \bibfield  {author} {\bibinfo {author} {\bibfnamefont {Mi~Young}\
  \bibnamefont {Choi}}, \bibinfo {author} {\bibfnamefont {Kwan~Woo}\
  \bibnamefont {Lee}}, \ and\ \bibinfo {author} {\bibfnamefont {Warren~E.}\
  \bibnamefont {Pickett}},\ }\bibfield  {title} {\enquote {\bibinfo {title}
  {{Role of 4f states in infinite-layer NdNiO2}},}\ }\href {\doibase
  10.1103/PhysRevB.101.020503} {\bibfield  {journal} {\bibinfo  {journal}
  {Physical Review B}\ }\textbf {\bibinfo {volume} {101}},\ \bibinfo {pages}
  {20503} (\bibinfo {year} {2020}{\natexlab{a}})}\BibitemShut {NoStop}%
\bibitem [{\citenamefont {Jiang}\ \emph {et~al.}(2019)\citenamefont {Jiang},
  \citenamefont {Si}, \citenamefont {Liao},\ and\ \citenamefont
  {Zhong}}]{Jiang2019}%
  \BibitemOpen
  \bibfield  {author} {\bibinfo {author} {\bibfnamefont {Peiheng}\ \bibnamefont
  {Jiang}}, \bibinfo {author} {\bibfnamefont {Liang}\ \bibnamefont {Si}},
  \bibinfo {author} {\bibfnamefont {Zhaoliang}\ \bibnamefont {Liao}}, \ and\
  \bibinfo {author} {\bibfnamefont {Zhicheng}\ \bibnamefont {Zhong}},\
  }\bibfield  {title} {\enquote {\bibinfo {title} {Electronic structure of
  rare-earth infinite-layer
  $r\mathrm{Ni}{\mathrm{o}}_{2}(r=\mathrm{La},\mathrm{Nd})$},}\ }\href
  {\doibase 10.1103/PhysRevB.100.201106} {\bibfield  {journal} {\bibinfo
  {journal} {Phys. Rev. B}\ }\textbf {\bibinfo {volume} {100}},\ \bibinfo
  {pages} {201106} (\bibinfo {year} {2019})}\BibitemShut {NoStop}%
\bibitem [{\citenamefont {Lechermann}(2020)}]{Lechermann2020}%
  \BibitemOpen
  \bibfield  {author} {\bibinfo {author} {\bibfnamefont {Frank}\ \bibnamefont
  {Lechermann}},\ }\bibfield  {title} {\enquote {\bibinfo {title} {{Late
  transition metal oxides with infinite-layer structure: Nickelates versus
  cuprates}},}\ }\href {\doibase 10.1103/PhysRevB.101.081110} {\bibfield
  {journal} {\bibinfo  {journal} {Physical Review B}\ }\textbf {\bibinfo
  {volume} {101}},\ \bibinfo {pages} {081110} (\bibinfo {year} {2020})},\
  \Eprint {http://arxiv.org/abs/1911.11521} {1911.11521} \BibitemShut {NoStop}%
\bibitem [{\citenamefont {Leonov}\ \emph {et~al.}(2020)\citenamefont {Leonov},
  \citenamefont {Skornyakov},\ and\ \citenamefont {Savrasov}}]{Leonov2020}%
  \BibitemOpen
  \bibfield  {author} {\bibinfo {author} {\bibfnamefont {I.}~\bibnamefont
  {Leonov}}, \bibinfo {author} {\bibfnamefont {S.~L.}\ \bibnamefont
  {Skornyakov}}, \ and\ \bibinfo {author} {\bibfnamefont {S.~Y.}\ \bibnamefont
  {Savrasov}},\ }\bibfield  {title} {\enquote {\bibinfo {title} {Lifshitz
  transition and frustration of magnetic moments in infinite-layer
  ${\mathrm{ndnio}}_{2}$ upon hole doping},}\ }\href {\doibase
  10.1103/PhysRevB.101.241108} {\bibfield  {journal} {\bibinfo  {journal}
  {Phys. Rev. B}\ }\textbf {\bibinfo {volume} {101}},\ \bibinfo {pages}
  {241108} (\bibinfo {year} {2020})}\BibitemShut {NoStop}%
\bibitem [{\citenamefont {Lee}\ and\ \citenamefont {Pickett}(2004)}]{Lee2004}%
  \BibitemOpen
  \bibfield  {author} {\bibinfo {author} {\bibfnamefont {K.-W.}\ \bibnamefont
  {Lee}}\ and\ \bibinfo {author} {\bibfnamefont {W.~E.}\ \bibnamefont
  {Pickett}},\ }\bibfield  {title} {\enquote {\bibinfo {title} {Infinite-layer
  $\mathrm{La}\mathrm{Ni}{\mathrm{o}}_{2}$: ${\mathrm{ni}}^{1+}$ is not
  ${\mathrm{cu}}^{2+}$},}\ }\href {\doibase 10.1103/PhysRevB.70.165109}
  {\bibfield  {journal} {\bibinfo  {journal} {Phys. Rev. B}\ }\textbf {\bibinfo
  {volume} {70}},\ \bibinfo {pages} {165109} (\bibinfo {year}
  {2004})}\BibitemShut {NoStop}%
\bibitem [{\citenamefont {Li}\ \emph {et~al.}(2020)\citenamefont {Li},
  \citenamefont {Wang}, \citenamefont {Lee}, \citenamefont {Harvey},
  \citenamefont {Osada}, \citenamefont {Goodge}, \citenamefont {Kourkoutis},\
  and\ \citenamefont {Hwang}}]{Zhang2020d}%
  \BibitemOpen
  \bibfield  {author} {\bibinfo {author} {\bibfnamefont {Danfeng}\ \bibnamefont
  {Li}}, \bibinfo {author} {\bibfnamefont {Bai~Yang}\ \bibnamefont {Wang}},
  \bibinfo {author} {\bibfnamefont {Kyuho}\ \bibnamefont {Lee}}, \bibinfo
  {author} {\bibfnamefont {Shannon~P.}\ \bibnamefont {Harvey}}, \bibinfo
  {author} {\bibfnamefont {Motoki}\ \bibnamefont {Osada}}, \bibinfo {author}
  {\bibfnamefont {Berit~H.}\ \bibnamefont {Goodge}}, \bibinfo {author}
  {\bibfnamefont {Lena~F.}\ \bibnamefont {Kourkoutis}}, \ and\ \bibinfo
  {author} {\bibfnamefont {Harold~Y.}\ \bibnamefont {Hwang}},\ }\bibfield
  {title} {\enquote {\bibinfo {title} {Superconducting dome in
  ${\mathrm{nd}}_{1\ensuremath{-}x}{\mathrm{sr}}_{x}{\mathrm{nio}}_{2}$
  infinite layer films},}\ }\href {\doibase 10.1103/PhysRevLett.125.027001}
  {\bibfield  {journal} {\bibinfo  {journal} {Phys. Rev. Lett.}\ }\textbf
  {\bibinfo {volume} {125}},\ \bibinfo {pages} {027001} (\bibinfo {year}
  {2020})}\BibitemShut {NoStop}%
\bibitem [{\citenamefont {Sakakibara}\ \emph {et~al.}(2020)\citenamefont
  {Sakakibara}, \citenamefont {Usui}, \citenamefont {Suzuki}, \citenamefont
  {Kotani}, \citenamefont {Aoki},\ and\ \citenamefont
  {Kuroki}}]{Sakakibara2019}%
  \BibitemOpen
  \bibfield  {author} {\bibinfo {author} {\bibfnamefont {Hirofumi}\
  \bibnamefont {Sakakibara}}, \bibinfo {author} {\bibfnamefont {Hidetomo}\
  \bibnamefont {Usui}}, \bibinfo {author} {\bibfnamefont {Katsuhiro}\
  \bibnamefont {Suzuki}}, \bibinfo {author} {\bibfnamefont {Takao}\
  \bibnamefont {Kotani}}, \bibinfo {author} {\bibfnamefont {Hideo}\
  \bibnamefont {Aoki}}, \ and\ \bibinfo {author} {\bibfnamefont {Kazuhiko}\
  \bibnamefont {Kuroki}},\ }\bibfield  {title} {\enquote {\bibinfo {title}
  {Model construction and a possibility of cupratelike pairing in a new
  ${d}^{9}$ nickelate superconductor
  $(\mathrm{Nd},\mathrm{Sr}){\mathrm{nio}}_{2}$},}\ }\href {\doibase
  10.1103/PhysRevLett.125.077003} {\bibfield  {journal} {\bibinfo  {journal}
  {Phys. Rev. Lett.}\ }\textbf {\bibinfo {volume} {125}},\ \bibinfo {pages}
  {077003} (\bibinfo {year} {2020})}\BibitemShut {NoStop}%
\bibitem [{\citenamefont {Botana}\ and\ \citenamefont
  {Norman}(2020)}]{Botana2020}%
  \BibitemOpen
  \bibfield  {author} {\bibinfo {author} {\bibfnamefont {A.~S.}\ \bibnamefont
  {Botana}}\ and\ \bibinfo {author} {\bibfnamefont {M.~R.}\ \bibnamefont
  {Norman}},\ }\bibfield  {title} {\enquote {\bibinfo {title} {Similarities and
  differences between ${\mathrm{lanio}}_{2}$ and ${\mathrm{cacuo}}_{2}$ and
  implications for superconductivity},}\ }\href {\doibase
  10.1103/PhysRevX.10.011024} {\bibfield  {journal} {\bibinfo  {journal} {Phys.
  Rev. X}\ }\textbf {\bibinfo {volume} {10}},\ \bibinfo {pages} {011024}
  (\bibinfo {year} {2020})}\BibitemShut {NoStop}%
\bibitem [{\citenamefont {Karp}\ \emph {et~al.}(2020)\citenamefont {Karp},
  \citenamefont {Botana}, \citenamefont {Norman}, \citenamefont {Park},
  \citenamefont {Zingl},\ and\ \citenamefont {Millis}}]{Karp2020}%
  \BibitemOpen
  \bibfield  {author} {\bibinfo {author} {\bibfnamefont {Jonathan}\
  \bibnamefont {Karp}}, \bibinfo {author} {\bibfnamefont {Antia~S.}\
  \bibnamefont {Botana}}, \bibinfo {author} {\bibfnamefont {Michael~R.}\
  \bibnamefont {Norman}}, \bibinfo {author} {\bibfnamefont {Hyowon}\
  \bibnamefont {Park}}, \bibinfo {author} {\bibfnamefont {Manuel}\ \bibnamefont
  {Zingl}}, \ and\ \bibinfo {author} {\bibfnamefont {Andrew}\ \bibnamefont
  {Millis}},\ }\bibfield  {title} {\enquote {\bibinfo {title} {Many-body
  electronic structure of ${\mathrm{ndnio}}_{2}$ and ${\mathrm{cacuo}}_{2}$},}\
  }\href {\doibase 10.1103/PhysRevX.10.021061} {\bibfield  {journal} {\bibinfo
  {journal} {Phys. Rev. X}\ }\textbf {\bibinfo {volume} {10}},\ \bibinfo
  {pages} {021061} (\bibinfo {year} {2020})}\BibitemShut {NoStop}%
\bibitem [{\citenamefont {Goodge}\ \emph {et~al.}(2021)\citenamefont {Goodge},
  \citenamefont {Li}, \citenamefont {Lee}, \citenamefont {Osada}, \citenamefont
  {Wang}, \citenamefont {Sawatzky}, \citenamefont {Hwang},\ and\ \citenamefont
  {Kourkoutis}}]{Goodge2021}%
  \BibitemOpen
  \bibfield  {author} {\bibinfo {author} {\bibfnamefont {Berit~H.}\
  \bibnamefont {Goodge}}, \bibinfo {author} {\bibfnamefont {Danfeng}\
  \bibnamefont {Li}}, \bibinfo {author} {\bibfnamefont {Kyuho}\ \bibnamefont
  {Lee}}, \bibinfo {author} {\bibfnamefont {Motoki}\ \bibnamefont {Osada}},
  \bibinfo {author} {\bibfnamefont {Bai~Yang}\ \bibnamefont {Wang}}, \bibinfo
  {author} {\bibfnamefont {George~A.}\ \bibnamefont {Sawatzky}}, \bibinfo
  {author} {\bibfnamefont {Harold~Y.}\ \bibnamefont {Hwang}}, \ and\ \bibinfo
  {author} {\bibfnamefont {Lena~F.}\ \bibnamefont {Kourkoutis}},\ }\bibfield
  {title} {\enquote {\bibinfo {title} {{Doping evolution of the Mott–Hubbard
  landscape in infinite-layer nickelates}},}\ }\href {\doibase
  10.1073/pnas.2007683118} {\bibfield  {journal} {\bibinfo  {journal}
  {Proceedings of the National Academy of Sciences}\ }\textbf {\bibinfo
  {volume} {118}},\ \bibinfo {pages} {e2007683118} (\bibinfo {year}
  {2021})}\BibitemShut {NoStop}%
\bibitem [{\citenamefont {Jiang}\ \emph {et~al.}(2020)\citenamefont {Jiang},
  \citenamefont {Berciu},\ and\ \citenamefont {Sawatzky}}]{Jiang2020}%
  \BibitemOpen
  \bibfield  {author} {\bibinfo {author} {\bibfnamefont {Mi}~\bibnamefont
  {Jiang}}, \bibinfo {author} {\bibfnamefont {Mona}\ \bibnamefont {Berciu}}, \
  and\ \bibinfo {author} {\bibfnamefont {George~A.}\ \bibnamefont {Sawatzky}},\
  }\bibfield  {title} {\enquote {\bibinfo {title} {{Critical Nature of the Ni
  Spin State in Doped ${\mathrm{NdNiO}}_{2}$}},}\ }\href {\doibase
  10.1103/PhysRevLett.124.207004} {\bibfield  {journal} {\bibinfo  {journal}
  {Phys. Rev. Lett.}\ }\textbf {\bibinfo {volume} {124}},\ \bibinfo {pages}
  {207004} (\bibinfo {year} {2020})}\BibitemShut {NoStop}%
\bibitem [{\citenamefont {Zhang}\ \emph {et~al.}(2021)\citenamefont {Zhang},
  \citenamefont {Lane}, \citenamefont {Singh}, \citenamefont {Nokelainen},
  \citenamefont {Barbiellini}, \citenamefont {Markiewicz}, \citenamefont
  {Bansil},\ and\ \citenamefont {Sun}}]{Zhang2021}%
  \BibitemOpen
  \bibfield  {author} {\bibinfo {author} {\bibfnamefont {Ruiqi}\ \bibnamefont
  {Zhang}}, \bibinfo {author} {\bibfnamefont {Christopher}\ \bibnamefont
  {Lane}}, \bibinfo {author} {\bibfnamefont {Bahadur}\ \bibnamefont {Singh}},
  \bibinfo {author} {\bibfnamefont {Johannes}\ \bibnamefont {Nokelainen}},
  \bibinfo {author} {\bibfnamefont {Bernardo}\ \bibnamefont {Barbiellini}},
  \bibinfo {author} {\bibfnamefont {Robert~S.}\ \bibnamefont {Markiewicz}},
  \bibinfo {author} {\bibfnamefont {Arun}\ \bibnamefont {Bansil}}, \ and\
  \bibinfo {author} {\bibfnamefont {Jianwei}\ \bibnamefont {Sun}},\ }\bibfield
  {title} {\enquote {\bibinfo {title} {{Magnetic and f-electron effects in
  LaNiO$_{2}$ and NdNiO$_{2}$ nickelates with cuprate-like 3$d_{x^2-y^2}$
  band}},}\ }\href {\doibase 10.1038/s42005-021-00621-4} {\bibfield  {journal}
  {\bibinfo  {journal} {Communications Physics}\ }\textbf {\bibinfo {volume}
  {4}},\ \bibinfo {pages} {118} (\bibinfo {year} {2021})}\BibitemShut {NoStop}%
\bibitem [{\citenamefont {{Zhang}}\ \emph {et~al.}(2022)\citenamefont
  {{Zhang}}, \citenamefont {{Lane}}, \citenamefont {{Nokelainen}},
  \citenamefont {{Singh}}, \citenamefont {{Barbiellini}}, \citenamefont
  {{Markiewicz}}, \citenamefont {{Bansil}},\ and\ \citenamefont
  {{Sun}}}]{2022arXiv220700184Z}%
  \BibitemOpen
  \bibfield  {author} {\bibinfo {author} {\bibfnamefont {Ruiqi}\ \bibnamefont
  {{Zhang}}}, \bibinfo {author} {\bibfnamefont {Christopher}\ \bibnamefont
  {{Lane}}}, \bibinfo {author} {\bibfnamefont {Johannes}\ \bibnamefont
  {{Nokelainen}}}, \bibinfo {author} {\bibfnamefont {Bahadur}\ \bibnamefont
  {{Singh}}}, \bibinfo {author} {\bibfnamefont {Bernardo}\ \bibnamefont
  {{Barbiellini}}}, \bibinfo {author} {\bibfnamefont {Robert~S.}\ \bibnamefont
  {{Markiewicz}}}, \bibinfo {author} {\bibfnamefont {Arun}\ \bibnamefont
  {{Bansil}}}, \ and\ \bibinfo {author} {\bibfnamefont {Jianwei}\ \bibnamefont
  {{Sun}}},\ }\bibfield  {title} {\enquote {\bibinfo {title} {{Peierls
  distortion driven multi-orbital origin of charge density waves in the undoped
  infinite-layer nickelate}},}\ }\href {\doibase 10.48550/arXiv.2207.00184}
  {\bibfield  {journal} {\bibinfo  {journal} {arXiv e-prints}\ } (\bibinfo
  {year} {2022}),\ 10.48550/arXiv.2207.00184}\BibitemShut {NoStop}%
\bibitem [{\citenamefont {Lane}\ \emph {et~al.}(2023)\citenamefont {Lane},
  \citenamefont {Zhang}, \citenamefont {Barbiellini}, \citenamefont
  {Markiewicz}, \citenamefont {Bansil}, \citenamefont {Sun},\ and\
  \citenamefont {Zhu}}]{Lane2023}%
  \BibitemOpen
  \bibfield  {author} {\bibinfo {author} {\bibfnamefont {Christopher}\
  \bibnamefont {Lane}}, \bibinfo {author} {\bibfnamefont {Ruiqi}\ \bibnamefont
  {Zhang}}, \bibinfo {author} {\bibfnamefont {Bernardo}\ \bibnamefont
  {Barbiellini}}, \bibinfo {author} {\bibfnamefont {Robert~S.}\ \bibnamefont
  {Markiewicz}}, \bibinfo {author} {\bibfnamefont {Arun}\ \bibnamefont
  {Bansil}}, \bibinfo {author} {\bibfnamefont {Jianwei}\ \bibnamefont {Sun}}, \
  and\ \bibinfo {author} {\bibfnamefont {Jian-Xin}\ \bibnamefont {Zhu}},\
  }\bibfield  {title} {\enquote {\bibinfo {title} {Competing incommensurate
  spin fluctuations and magnetic excitations in infinite-layer nickelate
  superconductors},}\ }\href {\doibase 10.1038/s42005-023-01213-0} {\bibfield
  {journal} {\bibinfo  {journal} {Communications Physics}\ }\textbf {\bibinfo
  {volume} {6}},\ \bibinfo {pages} {90} (\bibinfo {year} {2023})}\BibitemShut
  {NoStop}%
\bibitem [{\citenamefont {Rossi}\ \emph {et~al.}(2022)\citenamefont {Rossi},
  \citenamefont {Osada}, \citenamefont {Choi}, \citenamefont {Agrestini},
  \citenamefont {Jost}, \citenamefont {Lee}, \citenamefont {Lu}, \citenamefont
  {Wang}, \citenamefont {Lee}, \citenamefont {Nag}, \citenamefont {Chuang},
  \citenamefont {Kuo}, \citenamefont {Lee}, \citenamefont {Moritz},
  \citenamefont {Devereaux}, \citenamefont {Shen}, \citenamefont {Lee},
  \citenamefont {Zhou}, \citenamefont {Hwang},\ and\ \citenamefont
  {Lee}}]{2021arXiv211202484R}%
  \BibitemOpen
  \bibfield  {author} {\bibinfo {author} {\bibfnamefont {Matteo}\ \bibnamefont
  {Rossi}}, \bibinfo {author} {\bibfnamefont {Motoki}\ \bibnamefont {Osada}},
  \bibinfo {author} {\bibfnamefont {Jaewon}\ \bibnamefont {Choi}}, \bibinfo
  {author} {\bibfnamefont {Stefano}\ \bibnamefont {Agrestini}}, \bibinfo
  {author} {\bibfnamefont {Daniel}\ \bibnamefont {Jost}}, \bibinfo {author}
  {\bibfnamefont {Yonghun}\ \bibnamefont {Lee}}, \bibinfo {author}
  {\bibfnamefont {Haiyu}\ \bibnamefont {Lu}}, \bibinfo {author} {\bibfnamefont
  {Bai~Yang}\ \bibnamefont {Wang}}, \bibinfo {author} {\bibfnamefont {Kyuho}\
  \bibnamefont {Lee}}, \bibinfo {author} {\bibfnamefont {Abhishek}\
  \bibnamefont {Nag}}, \bibinfo {author} {\bibfnamefont {Yi-De}\ \bibnamefont
  {Chuang}}, \bibinfo {author} {\bibfnamefont {Cheng-Tai}\ \bibnamefont {Kuo}},
  \bibinfo {author} {\bibfnamefont {Sang-Jun}\ \bibnamefont {Lee}}, \bibinfo
  {author} {\bibfnamefont {Brian}\ \bibnamefont {Moritz}}, \bibinfo {author}
  {\bibfnamefont {Thomas~P.}\ \bibnamefont {Devereaux}}, \bibinfo {author}
  {\bibfnamefont {Zhi-Xun}\ \bibnamefont {Shen}}, \bibinfo {author}
  {\bibfnamefont {Jun-Sik}\ \bibnamefont {Lee}}, \bibinfo {author}
  {\bibfnamefont {Ke-Jin}\ \bibnamefont {Zhou}}, \bibinfo {author}
  {\bibfnamefont {Harold~Y.}\ \bibnamefont {Hwang}}, \ and\ \bibinfo {author}
  {\bibfnamefont {Wei-Sheng}\ \bibnamefont {Lee}},\ }\bibfield  {title}
  {\enquote {\bibinfo {title} {{A broken translational symmetry state in an
  infinite-layer nickelate}},}\ }\href {\doibase 10.1038/s41567-022-01660-6}
  {\bibfield  {journal} {\bibinfo  {journal} {Nat. Phys.}\ }\textbf {\bibinfo
  {volume} {18}},\ \bibinfo {pages} {869--873} (\bibinfo {year}
  {2022})}\BibitemShut {NoStop}%
\bibitem [{\citenamefont {Krieger}\ \emph {et~al.}(2022)\citenamefont
  {Krieger}, \citenamefont {Martinelli}, \citenamefont {Zeng}, \citenamefont
  {Chow}, \citenamefont {Kummer}, \citenamefont {Arpaia}, \citenamefont
  {Moretti~Sala}, \citenamefont {Brookes}, \citenamefont {Ariando},
  \citenamefont {Viart}, \citenamefont {Salluzzo}, \citenamefont
  {Ghiringhelli},\ and\ \citenamefont {Preziosi}}]{Krieger2021}%
  \BibitemOpen
  \bibfield  {author} {\bibinfo {author} {\bibfnamefont {G.}~\bibnamefont
  {Krieger}}, \bibinfo {author} {\bibfnamefont {L.}~\bibnamefont {Martinelli}},
  \bibinfo {author} {\bibfnamefont {S.}~\bibnamefont {Zeng}}, \bibinfo {author}
  {\bibfnamefont {L.~E.}\ \bibnamefont {Chow}}, \bibinfo {author}
  {\bibfnamefont {K.}~\bibnamefont {Kummer}}, \bibinfo {author} {\bibfnamefont
  {R.}~\bibnamefont {Arpaia}}, \bibinfo {author} {\bibfnamefont
  {M.}~\bibnamefont {Moretti~Sala}}, \bibinfo {author} {\bibfnamefont {N.~B.}\
  \bibnamefont {Brookes}}, \bibinfo {author} {\bibfnamefont {A.}~\bibnamefont
  {Ariando}}, \bibinfo {author} {\bibfnamefont {N.}~\bibnamefont {Viart}},
  \bibinfo {author} {\bibfnamefont {M.}~\bibnamefont {Salluzzo}}, \bibinfo
  {author} {\bibfnamefont {G.}~\bibnamefont {Ghiringhelli}}, \ and\ \bibinfo
  {author} {\bibfnamefont {D.}~\bibnamefont {Preziosi}},\ }\bibfield  {title}
  {\enquote {\bibinfo {title} {Charge and spin order dichotomy in
  ${\mathrm{ndnio}}_{2}$ driven by the capping layer},}\ }\href {\doibase
  10.1103/PhysRevLett.129.027002} {\bibfield  {journal} {\bibinfo  {journal}
  {Phys. Rev. Lett.}\ }\textbf {\bibinfo {volume} {129}},\ \bibinfo {pages}
  {027002} (\bibinfo {year} {2022})}\BibitemShut {NoStop}%
\bibitem [{\citenamefont {Tam}\ \emph {et~al.}(2022)\citenamefont {Tam},
  \citenamefont {Choi}, \citenamefont {Ding}, \citenamefont {Agrestini},
  \citenamefont {Nag}, \citenamefont {Wu}, \citenamefont {Huang}, \citenamefont
  {Luo}, \citenamefont {Gao}, \citenamefont {Garc{\'{i}}a-Fern{\'{a}}ndez},
  \citenamefont {Qiao},\ and\ \citenamefont {Zhou}}]{Tam2021}%
  \BibitemOpen
  \bibfield  {author} {\bibinfo {author} {\bibfnamefont {Charles~C.}\
  \bibnamefont {Tam}}, \bibinfo {author} {\bibfnamefont {Jaewon}\ \bibnamefont
  {Choi}}, \bibinfo {author} {\bibfnamefont {Xiang}\ \bibnamefont {Ding}},
  \bibinfo {author} {\bibfnamefont {Stefano}\ \bibnamefont {Agrestini}},
  \bibinfo {author} {\bibfnamefont {Abhishek}\ \bibnamefont {Nag}}, \bibinfo
  {author} {\bibfnamefont {Mei}\ \bibnamefont {Wu}}, \bibinfo {author}
  {\bibfnamefont {Bing}\ \bibnamefont {Huang}}, \bibinfo {author}
  {\bibfnamefont {Huiqian}\ \bibnamefont {Luo}}, \bibinfo {author}
  {\bibfnamefont {Peng}\ \bibnamefont {Gao}}, \bibinfo {author} {\bibfnamefont
  {Mirian}\ \bibnamefont {Garc{\'{i}}a-Fern{\'{a}}ndez}}, \bibinfo {author}
  {\bibfnamefont {Liang}\ \bibnamefont {Qiao}}, \ and\ \bibinfo {author}
  {\bibfnamefont {Ke-Jin}\ \bibnamefont {Zhou}},\ }\bibfield  {title} {\enquote
  {\bibinfo {title} {{Charge density waves in infinite-layer NdNiO$_{2}$
  nickelates}},}\ }\href {\doibase 10.1038/s41563-022-01330-1} {\bibfield
  {journal} {\bibinfo  {journal} {Nat. Mater.}\ }\textbf {\bibinfo {volume}
  {21}},\ \bibinfo {pages} {1116--1120} (\bibinfo {year} {2022})},\ \Eprint
  {http://arxiv.org/abs/2112.04440} {2112.04440} \BibitemShut {NoStop}%
\bibitem [{\citenamefont {Osada}\ \emph
  {et~al.}(2021{\natexlab{b}})\citenamefont {Osada}, \citenamefont {Wang},
  \citenamefont {Goodge}, \citenamefont {Harvey}, \citenamefont {Lee},
  \citenamefont {Li}, \citenamefont {Kourkoutis},\ and\ \citenamefont
  {Hwang}}]{OsadaAM2021}%
  \BibitemOpen
  \bibfield  {author} {\bibinfo {author} {\bibfnamefont {Motoki}\ \bibnamefont
  {Osada}}, \bibinfo {author} {\bibfnamefont {Bai~Yang}\ \bibnamefont {Wang}},
  \bibinfo {author} {\bibfnamefont {Berit~H.}\ \bibnamefont {Goodge}}, \bibinfo
  {author} {\bibfnamefont {Shannon~P.}\ \bibnamefont {Harvey}}, \bibinfo
  {author} {\bibfnamefont {Kyuho}\ \bibnamefont {Lee}}, \bibinfo {author}
  {\bibfnamefont {Danfeng}\ \bibnamefont {Li}}, \bibinfo {author}
  {\bibfnamefont {Lena~F.}\ \bibnamefont {Kourkoutis}}, \ and\ \bibinfo
  {author} {\bibfnamefont {Harold~Y.}\ \bibnamefont {Hwang}},\ }\bibfield
  {title} {\enquote {\bibinfo {title} {Nickelate superconductivity without
  rare-earth magnetism: (la,sr)nio2},}\ }\href {\doibase
  https://doi.org/10.1002/adma.202104083} {\bibfield  {journal} {\bibinfo
  {journal} {Advanced Materials}\ }\textbf {\bibinfo {volume} {33}},\ \bibinfo
  {pages} {2104083} (\bibinfo {year} {2021}{\natexlab{b}})}\BibitemShut
  {NoStop}%
\bibitem [{\citenamefont {Fowlie}\ \emph {et~al.}(2022)\citenamefont {Fowlie},
  \citenamefont {Hadjimichael}, \citenamefont {Martins}, \citenamefont {Li},
  \citenamefont {Osada}, \citenamefont {Wang}, \citenamefont {Lee},
  \citenamefont {Lee}, \citenamefont {Salman}, \citenamefont {Prokscha},
  \citenamefont {Triscone}, \citenamefont {Hwang},\ and\ \citenamefont
  {Suter}}]{Fowlie2022}%
  \BibitemOpen
  \bibfield  {author} {\bibinfo {author} {\bibfnamefont {Jennifer}\
  \bibnamefont {Fowlie}}, \bibinfo {author} {\bibfnamefont {Marios}\
  \bibnamefont {Hadjimichael}}, \bibinfo {author} {\bibfnamefont {Maria~M.}\
  \bibnamefont {Martins}}, \bibinfo {author} {\bibfnamefont {Danfeng}\
  \bibnamefont {Li}}, \bibinfo {author} {\bibfnamefont {Motoki}\ \bibnamefont
  {Osada}}, \bibinfo {author} {\bibfnamefont {Bai~Yang}\ \bibnamefont {Wang}},
  \bibinfo {author} {\bibfnamefont {Kyuho}\ \bibnamefont {Lee}}, \bibinfo
  {author} {\bibfnamefont {Yonghun}\ \bibnamefont {Lee}}, \bibinfo {author}
  {\bibfnamefont {Zaher}\ \bibnamefont {Salman}}, \bibinfo {author}
  {\bibfnamefont {Thomas}\ \bibnamefont {Prokscha}}, \bibinfo {author}
  {\bibfnamefont {Jean-Marc}\ \bibnamefont {Triscone}}, \bibinfo {author}
  {\bibfnamefont {Harold~Y.}\ \bibnamefont {Hwang}}, \ and\ \bibinfo {author}
  {\bibfnamefont {Andreas}\ \bibnamefont {Suter}},\ }\bibfield  {title}
  {\enquote {\bibinfo {title} {{Intrinsic magnetism in superconducting
  infinite-layer nickelates}},}\ }\href {\doibase 10.1038/s41567-022-01684-y}
  {\bibfield  {journal} {\bibinfo  {journal} {Nat. Phys.}\ }\textbf {\bibinfo
  {volume} {18}},\ \bibinfo {pages} {1043--1047} (\bibinfo {year}
  {2022})}\BibitemShut {NoStop}%
\bibitem [{\citenamefont {Lu}\ \emph {et~al.}(2021)\citenamefont {Lu},
  \citenamefont {Rossi}, \citenamefont {Nag}, \citenamefont {Osada},
  \citenamefont {Li}, \citenamefont {Lee}, \citenamefont {Wang}, \citenamefont
  {Garcia-Fernandez}, \citenamefont {Agrestini}, \citenamefont {Shen},
  \citenamefont {Been}, \citenamefont {Moritz}, \citenamefont {Devereaux},
  \citenamefont {Zaanen}, \citenamefont {Hwang}, \citenamefont {Zhou},\ and\
  \citenamefont {Lee}}]{Lu2021science}%
  \BibitemOpen
  \bibfield  {author} {\bibinfo {author} {\bibfnamefont {H.}~\bibnamefont
  {Lu}}, \bibinfo {author} {\bibfnamefont {M.}~\bibnamefont {Rossi}}, \bibinfo
  {author} {\bibfnamefont {A.}~\bibnamefont {Nag}}, \bibinfo {author}
  {\bibfnamefont {M.}~\bibnamefont {Osada}}, \bibinfo {author} {\bibfnamefont
  {D.~F.}\ \bibnamefont {Li}}, \bibinfo {author} {\bibfnamefont
  {K.}~\bibnamefont {Lee}}, \bibinfo {author} {\bibfnamefont {B.~Y.}\
  \bibnamefont {Wang}}, \bibinfo {author} {\bibfnamefont {M.}~\bibnamefont
  {Garcia-Fernandez}}, \bibinfo {author} {\bibfnamefont {S.}~\bibnamefont
  {Agrestini}}, \bibinfo {author} {\bibfnamefont {Z.~X.}\ \bibnamefont {Shen}},
  \bibinfo {author} {\bibfnamefont {E.~M.}\ \bibnamefont {Been}}, \bibinfo
  {author} {\bibfnamefont {B.}~\bibnamefont {Moritz}}, \bibinfo {author}
  {\bibfnamefont {T.~P.}\ \bibnamefont {Devereaux}}, \bibinfo {author}
  {\bibfnamefont {J.}~\bibnamefont {Zaanen}}, \bibinfo {author} {\bibfnamefont
  {H.~Y.}\ \bibnamefont {Hwang}}, \bibinfo {author} {\bibfnamefont {Ke-Jin}\
  \bibnamefont {Zhou}}, \ and\ \bibinfo {author} {\bibfnamefont {W.~S.}\
  \bibnamefont {Lee}},\ }\bibfield  {title} {\enquote {\bibinfo {title}
  {Magnetic excitations in infinite-layer nickelates},}\ }\href {\doibase
  10.1126/science.abd7726} {\bibfield  {journal} {\bibinfo  {journal}
  {Science}\ }\textbf {\bibinfo {volume} {373}},\ \bibinfo {pages} {213--216}
  (\bibinfo {year} {2021})}\BibitemShut {NoStop}%
\bibitem [{\citenamefont {{Ortiz}}\ \emph {et~al.}(2021)\citenamefont
  {{Ortiz}}, \citenamefont {{Puphal}}, \citenamefont {{Klett}}, \citenamefont
  {{Hotz}}, \citenamefont {{Kremer}}, \citenamefont {{Trepka}}, \citenamefont
  {{Hemmida}}, \citenamefont {{Krug von Nidda}}, \citenamefont {{Isobe}},
  \citenamefont {{Khasanov}}, \citenamefont {{Luetkens}}, \citenamefont
  {{Hansmann}}, \citenamefont {{Keimer}}, \citenamefont {{Sch{\"a}fer}},\ and\
  \citenamefont {{Hepting}}}]{2021arXiv211113668O}%
  \BibitemOpen
  \bibfield  {author} {\bibinfo {author} {\bibfnamefont {R.~A.}\ \bibnamefont
  {{Ortiz}}}, \bibinfo {author} {\bibfnamefont {P.}~\bibnamefont {{Puphal}}},
  \bibinfo {author} {\bibfnamefont {M.}~\bibnamefont {{Klett}}}, \bibinfo
  {author} {\bibfnamefont {F.}~\bibnamefont {{Hotz}}}, \bibinfo {author}
  {\bibfnamefont {R.~K.}\ \bibnamefont {{Kremer}}}, \bibinfo {author}
  {\bibfnamefont {H.}~\bibnamefont {{Trepka}}}, \bibinfo {author}
  {\bibfnamefont {M.}~\bibnamefont {{Hemmida}}}, \bibinfo {author}
  {\bibfnamefont {H.~A.}\ \bibnamefont {{Krug von Nidda}}}, \bibinfo {author}
  {\bibfnamefont {M.}~\bibnamefont {{Isobe}}}, \bibinfo {author} {\bibfnamefont
  {R.}~\bibnamefont {{Khasanov}}}, \bibinfo {author} {\bibfnamefont
  {H.}~\bibnamefont {{Luetkens}}}, \bibinfo {author} {\bibfnamefont
  {P.}~\bibnamefont {{Hansmann}}}, \bibinfo {author} {\bibfnamefont
  {B.}~\bibnamefont {{Keimer}}}, \bibinfo {author} {\bibfnamefont
  {T.}~\bibnamefont {{Sch{\"a}fer}}}, \ and\ \bibinfo {author} {\bibfnamefont
  {M.}~\bibnamefont {{Hepting}}},\ }\bibfield  {title} {\enquote {\bibinfo
  {title} {{Magnetic correlations in infinite-layer nickelates: an experimental
  and theoretical multi-method study}},}\ }\href@noop {} {\bibfield  {journal}
  {\bibinfo  {journal} {arXiv e-prints}\ ,\ \bibinfo {eid} {arXiv:2111.13668}}
  (\bibinfo {year} {2021})},\ \Eprint {http://arxiv.org/abs/2111.13668}
  {arXiv:2111.13668 [cond-mat.str-el]} \BibitemShut {NoStop}%
\bibitem [{\citenamefont {Gu}\ \emph {et~al.}(2020)\citenamefont {Gu},
  \citenamefont {Zhu}, \citenamefont {Wang}, \citenamefont {Hu},\ and\
  \citenamefont {Chen}}]{Gu2020}%
  \BibitemOpen
  \bibfield  {author} {\bibinfo {author} {\bibfnamefont {Yuhao}\ \bibnamefont
  {Gu}}, \bibinfo {author} {\bibfnamefont {Sichen}\ \bibnamefont {Zhu}},
  \bibinfo {author} {\bibfnamefont {Xiaoxuan}\ \bibnamefont {Wang}}, \bibinfo
  {author} {\bibfnamefont {Jiangping}\ \bibnamefont {Hu}}, \ and\ \bibinfo
  {author} {\bibfnamefont {Hanghui}\ \bibnamefont {Chen}},\ }\bibfield  {title}
  {\enquote {\bibinfo {title} {{A substantial hybridization between correlated
  Ni-d orbital and itinerant electrons in infinite-layer nickelates}},}\ }\href
  {\doibase 10.1038/s42005-020-0347-x} {\bibfield  {journal} {\bibinfo
  {journal} {Communications Physics}\ }\textbf {\bibinfo {volume} {3}},\
  \bibinfo {pages} {84} (\bibinfo {year} {2020})}\BibitemShut {NoStop}%
\bibitem [{\citenamefont {Kitatani}\ \emph {et~al.}(2020)\citenamefont
  {Kitatani}, \citenamefont {Si}, \citenamefont {Janson}, \citenamefont
  {Arita}, \citenamefont {Zhong},\ and\ \citenamefont {Held}}]{Kitatani2020}%
  \BibitemOpen
  \bibfield  {author} {\bibinfo {author} {\bibfnamefont {Motoharu}\
  \bibnamefont {Kitatani}}, \bibinfo {author} {\bibfnamefont {Liang}\
  \bibnamefont {Si}}, \bibinfo {author} {\bibfnamefont {Oleg}\ \bibnamefont
  {Janson}}, \bibinfo {author} {\bibfnamefont {Ryotaro}\ \bibnamefont {Arita}},
  \bibinfo {author} {\bibfnamefont {Zhicheng}\ \bibnamefont {Zhong}}, \ and\
  \bibinfo {author} {\bibfnamefont {Karsten}\ \bibnamefont {Held}},\ }\bibfield
   {title} {\enquote {\bibinfo {title} {{Nickelate superconductors—a
  renaissance of the one-band Hubbard model}},}\ }\href {\doibase
  10.1038/s41535-020-00260-y} {\bibfield  {journal} {\bibinfo  {journal} {npj
  Quantum Materials}\ }\textbf {\bibinfo {volume} {5}},\ \bibinfo {pages} {59}
  (\bibinfo {year} {2020})}\BibitemShut {NoStop}%
\bibitem [{\citenamefont {{Chow}}\ \emph {et~al.}(2022)\citenamefont {{Chow}},
  \citenamefont {{Kunniniyil Sudheesh}}, \citenamefont {{Luo}}, \citenamefont
  {{Nandi}}, \citenamefont {{Heil}}, \citenamefont {{Deuschle}}, \citenamefont
  {{Zeng}}, \citenamefont {{Zhang}}, \citenamefont {{Prakash}}, \citenamefont
  {{Du}}, \citenamefont {{Lim}}, \citenamefont {{van Aken}}, \citenamefont
  {{Chia}},\ and\ \citenamefont {{Ariando}}}]{2022arXiv220110038C}%
  \BibitemOpen
  \bibfield  {author} {\bibinfo {author} {\bibfnamefont {L.~E.}\ \bibnamefont
  {{Chow}}}, \bibinfo {author} {\bibfnamefont {S.}~\bibnamefont {{Kunniniyil
  Sudheesh}}}, \bibinfo {author} {\bibfnamefont {Z.~Y.}\ \bibnamefont {{Luo}}},
  \bibinfo {author} {\bibfnamefont {P.}~\bibnamefont {{Nandi}}}, \bibinfo
  {author} {\bibfnamefont {T.}~\bibnamefont {{Heil}}}, \bibinfo {author}
  {\bibfnamefont {J.}~\bibnamefont {{Deuschle}}}, \bibinfo {author}
  {\bibfnamefont {S.~W.}\ \bibnamefont {{Zeng}}}, \bibinfo {author}
  {\bibfnamefont {Z.~T.}\ \bibnamefont {{Zhang}}}, \bibinfo {author}
  {\bibfnamefont {S.}~\bibnamefont {{Prakash}}}, \bibinfo {author}
  {\bibfnamefont {X.~M.}\ \bibnamefont {{Du}}}, \bibinfo {author}
  {\bibfnamefont {Z.~S.}\ \bibnamefont {{Lim}}}, \bibinfo {author}
  {\bibfnamefont {Peter~A.}\ \bibnamefont {{van Aken}}}, \bibinfo {author}
  {\bibfnamefont {Elbert E.~M.}\ \bibnamefont {{Chia}}}, \ and\ \bibinfo
  {author} {\bibfnamefont {A.}~\bibnamefont {{Ariando}}},\ }\bibfield  {title}
  {\enquote {\bibinfo {title} {{Pairing symmetry in infinite-layer nickelate
  superconductor}},}\ }\href {\doibase 10.48550/arXiv.2201.10038} {\bibfield
  {journal} {\bibinfo  {journal} {arXiv e-prints}\ ,\ \bibinfo {eid}
  {arXiv:2201.10038}} (\bibinfo {year} {2022})},\ \Eprint
  {http://arxiv.org/abs/2201.10038} {arXiv:2201.10038 [cond-mat.supr-con]}
  \BibitemShut {NoStop}%
\bibitem [{\citenamefont {Kreisel}\ \emph {et~al.}(2022)\citenamefont
  {Kreisel}, \citenamefont {Andersen}, \citenamefont {R\o{}mer}, \citenamefont
  {Eremin},\ and\ \citenamefont {Lechermann}}]{Kreisel2022PRL}%
  \BibitemOpen
  \bibfield  {author} {\bibinfo {author} {\bibfnamefont {Andreas}\ \bibnamefont
  {Kreisel}}, \bibinfo {author} {\bibfnamefont {Brian~M.}\ \bibnamefont
  {Andersen}}, \bibinfo {author} {\bibfnamefont {Astrid~T.}\ \bibnamefont
  {R\o{}mer}}, \bibinfo {author} {\bibfnamefont {Ilya~M.}\ \bibnamefont
  {Eremin}}, \ and\ \bibinfo {author} {\bibfnamefont {Frank}\ \bibnamefont
  {Lechermann}},\ }\bibfield  {title} {\enquote {\bibinfo {title}
  {Superconducting instabilities in strongly correlated infinite-layer
  nickelates},}\ }\href {\doibase 10.1103/PhysRevLett.129.077002} {\bibfield
  {journal} {\bibinfo  {journal} {Phys. Rev. Lett.}\ }\textbf {\bibinfo
  {volume} {129}},\ \bibinfo {pages} {077002} (\bibinfo {year}
  {2022})}\BibitemShut {NoStop}%
\bibitem [{\citenamefont {Gao}\ \emph {et~al.}(2020)\citenamefont {Gao},
  \citenamefont {Peng}, \citenamefont {Wang}, \citenamefont {Fang},\ and\
  \citenamefont {Weng}}]{GaoNSR2020}%
  \BibitemOpen
  \bibfield  {author} {\bibinfo {author} {\bibfnamefont {Jiacheng}\
  \bibnamefont {Gao}}, \bibinfo {author} {\bibfnamefont {Shiyu}\ \bibnamefont
  {Peng}}, \bibinfo {author} {\bibfnamefont {Zhijun}\ \bibnamefont {Wang}},
  \bibinfo {author} {\bibfnamefont {Chen}\ \bibnamefont {Fang}}, \ and\
  \bibinfo {author} {\bibfnamefont {Hongming}\ \bibnamefont {Weng}},\
  }\bibfield  {title} {\enquote {\bibinfo {title} {{Electronic structures and
  topological properties in nickelates Lnn+1NinO2n+2}},}\ }\href
  {https://doi.org/10.1093/nsr/nwaa218} {\bibfield  {journal} {\bibinfo
  {journal} {National Science Review}\ }\textbf {\bibinfo {volume} {8}}
  (\bibinfo {year} {2020})}\BibitemShut {NoStop}%
\bibitem [{\citenamefont {Choi}\ \emph
  {et~al.}(2020{\natexlab{b}})\citenamefont {Choi}, \citenamefont {Pickett},\
  and\ \citenamefont {Lee}}]{ChoiPRR2020}%
  \BibitemOpen
  \bibfield  {author} {\bibinfo {author} {\bibfnamefont {Mi-Young}\
  \bibnamefont {Choi}}, \bibinfo {author} {\bibfnamefont {Warren~E.}\
  \bibnamefont {Pickett}}, \ and\ \bibinfo {author} {\bibfnamefont {Kwan-Woo}\
  \bibnamefont {Lee}},\ }\bibfield  {title} {\enquote {\bibinfo {title}
  {Fluctuation-frustrated flat band instabilities in ${\mathrm{ndnio}}_{2}$},}\
  }\href {\doibase 10.1103/PhysRevResearch.2.033445} {\bibfield  {journal}
  {\bibinfo  {journal} {Phys. Rev. Research}\ }\textbf {\bibinfo {volume}
  {2}},\ \bibinfo {pages} {033445} (\bibinfo {year}
  {2020}{\natexlab{b}})}\BibitemShut {NoStop}%
\bibitem [{\citenamefont {Kresse}\ and\ \citenamefont
  {Joubert}(1999)}]{Kresse1999}%
  \BibitemOpen
  \bibfield  {author} {\bibinfo {author} {\bibfnamefont {G.}~\bibnamefont
  {Kresse}}\ and\ \bibinfo {author} {\bibfnamefont {D.}~\bibnamefont
  {Joubert}},\ }\bibfield  {title} {\enquote {\bibinfo {title} {{From ultrasoft
  pseudopotentials to the projector augmented-wave method}},}\ }\href {\doibase
  10.1103/PhysRevB.59.1758} {\bibfield  {journal} {\bibinfo  {journal}
  {Physical Review B}\ }\textbf {\bibinfo {volume} {59}},\ \bibinfo {pages}
  {1758--1775} (\bibinfo {year} {1999})}\BibitemShut {NoStop}%
\bibitem [{\citenamefont {Kresse}\ and\ \citenamefont
  {Hafner}(1993)}]{Kresse1993}%
  \BibitemOpen
  \bibfield  {author} {\bibinfo {author} {\bibfnamefont {G.}~\bibnamefont
  {Kresse}}\ and\ \bibinfo {author} {\bibfnamefont {J.}~\bibnamefont
  {Hafner}},\ }\bibfield  {title} {\enquote {\bibinfo {title} {{Ab initio
  molecular dynamics for open-shell transition metals}},}\ }\href {\doibase
  10.1103/PhysRevB.48.13115} {\bibfield  {journal} {\bibinfo  {journal}
  {Physical Review B}\ }\textbf {\bibinfo {volume} {48}},\ \bibinfo {pages}
  {13115--13118} (\bibinfo {year} {1993})}\BibitemShut {NoStop}%
\bibitem [{\citenamefont {Kresse}\ and\ \citenamefont
  {Furthm{\"{u}}ller}(1996)}]{Kresse1996}%
  \BibitemOpen
  \bibfield  {author} {\bibinfo {author} {\bibfnamefont {G}~\bibnamefont
  {Kresse}}\ and\ \bibinfo {author} {\bibfnamefont {J.}~\bibnamefont
  {Furthm{\"{u}}ller}},\ }\bibfield  {title} {\enquote {\bibinfo {title}
  {{Efficient iterative schemes for ab initio total-energy calculations using a
  plane-wave basis set}},}\ }\href {\doibase 10.1103/PhysRevB.54.11169}
  {\bibfield  {journal} {\bibinfo  {journal} {Physical Review B}\ }\textbf
  {\bibinfo {volume} {54}},\ \bibinfo {pages} {11169--11186} (\bibinfo {year}
  {1996})}\BibitemShut {NoStop}%
\bibitem [{\citenamefont {Sun}\ \emph {et~al.}(2015)\citenamefont {Sun},
  \citenamefont {Ruzsinszky},\ and\ \citenamefont {Perdew}}]{Sun2015}%
  \BibitemOpen
  \bibfield  {author} {\bibinfo {author} {\bibfnamefont {Jianwei}\ \bibnamefont
  {Sun}}, \bibinfo {author} {\bibfnamefont {Adrienn}\ \bibnamefont
  {Ruzsinszky}}, \ and\ \bibinfo {author} {\bibfnamefont {Johnp}\ \bibnamefont
  {Perdew}},\ }\bibfield  {title} {\enquote {\bibinfo {title} {{Strongly
  Constrained and Appropriately Normed Semilocal Density Functional}},}\ }\href
  {\doibase 10.1103/PhysRevLett.115.036402} {\bibfield  {journal} {\bibinfo
  {journal} {Physical Review Letters}\ }\textbf {\bibinfo {volume} {115}},\
  \bibinfo {pages} {036402} (\bibinfo {year} {2015})}\BibitemShut {NoStop}%
\bibitem [{\citenamefont {Zhang}\ \emph
  {et~al.}(2020{\natexlab{a}})\citenamefont {Zhang}, \citenamefont {Lane},
  \citenamefont {Furness}, \citenamefont {Barbiellini}, \citenamefont {Perdew},
  \citenamefont {Markiewicz}, \citenamefont {Bansil},\ and\ \citenamefont
  {Sun}}]{Zhang2020}%
  \BibitemOpen
  \bibfield  {author} {\bibinfo {author} {\bibfnamefont {Yubo}\ \bibnamefont
  {Zhang}}, \bibinfo {author} {\bibfnamefont {Christopher}\ \bibnamefont
  {Lane}}, \bibinfo {author} {\bibfnamefont {James~W.}\ \bibnamefont
  {Furness}}, \bibinfo {author} {\bibfnamefont {Bernardo}\ \bibnamefont
  {Barbiellini}}, \bibinfo {author} {\bibfnamefont {John~P.}\ \bibnamefont
  {Perdew}}, \bibinfo {author} {\bibfnamefont {Robert~S.}\ \bibnamefont
  {Markiewicz}}, \bibinfo {author} {\bibfnamefont {Arun}\ \bibnamefont
  {Bansil}}, \ and\ \bibinfo {author} {\bibfnamefont {Jianwei}\ \bibnamefont
  {Sun}},\ }\bibfield  {title} {\enquote {\bibinfo {title} {{Competing stripe
  and magnetic phases in the cuprates from first principles}},}\ }\href
  {\doibase 10.1073/pnas.1910411116} {\bibfield  {journal} {\bibinfo  {journal}
  {Proceedings of the National Academy of Sciences}\ }\textbf {\bibinfo
  {volume} {117}},\ \bibinfo {pages} {68--72} (\bibinfo {year}
  {2020}{\natexlab{a}})}\BibitemShut {NoStop}%
\bibitem [{\citenamefont {Furness}\ \emph {et~al.}(2018)\citenamefont
  {Furness}, \citenamefont {Zhang}, \citenamefont {Lane}, \citenamefont {Buda},
  \citenamefont {Barbiellini}, \citenamefont {Markiewicz}, \citenamefont
  {Bansil},\ and\ \citenamefont {Sun}}]{Furness2018}%
  \BibitemOpen
  \bibfield  {author} {\bibinfo {author} {\bibfnamefont {James~W.}\
  \bibnamefont {Furness}}, \bibinfo {author} {\bibfnamefont {Yubo}\
  \bibnamefont {Zhang}}, \bibinfo {author} {\bibfnamefont {Christopher}\
  \bibnamefont {Lane}}, \bibinfo {author} {\bibfnamefont {Ioana~Gianina}\
  \bibnamefont {Buda}}, \bibinfo {author} {\bibfnamefont {Bernardo}\
  \bibnamefont {Barbiellini}}, \bibinfo {author} {\bibfnamefont {Robert~S.}\
  \bibnamefont {Markiewicz}}, \bibinfo {author} {\bibfnamefont {Arun}\
  \bibnamefont {Bansil}}, \ and\ \bibinfo {author} {\bibfnamefont {Jianwei}\
  \bibnamefont {Sun}},\ }\bibfield  {title} {\enquote {\bibinfo {title} {{An
  accurate first-principles treatment of doping-dependent electronic structure
  of high-temperature cuprate superconductors}},}\ }\href {\doibase
  10.1038/s42005-018-0009-4} {\bibfield  {journal} {\bibinfo  {journal}
  {Communications Physics}\ }\textbf {\bibinfo {volume} {1}},\ \bibinfo {pages}
  {11} (\bibinfo {year} {2018})}\BibitemShut {NoStop}%
\bibitem [{\citenamefont {Lane}\ \emph {et~al.}(2018)\citenamefont {Lane},
  \citenamefont {Furness}, \citenamefont {Buda}, \citenamefont {Zhang},
  \citenamefont {Markiewicz}, \citenamefont {Barbiellini}, \citenamefont
  {Sun},\ and\ \citenamefont {Bansil}}]{Lane2018}%
  \BibitemOpen
  \bibfield  {author} {\bibinfo {author} {\bibfnamefont {Christopher}\
  \bibnamefont {Lane}}, \bibinfo {author} {\bibfnamefont {James~W.}\
  \bibnamefont {Furness}}, \bibinfo {author} {\bibfnamefont {Ioana~Gianina}\
  \bibnamefont {Buda}}, \bibinfo {author} {\bibfnamefont {Yubo}\ \bibnamefont
  {Zhang}}, \bibinfo {author} {\bibfnamefont {Robert~S.}\ \bibnamefont
  {Markiewicz}}, \bibinfo {author} {\bibfnamefont {Bernardo}\ \bibnamefont
  {Barbiellini}}, \bibinfo {author} {\bibfnamefont {Jianwei}\ \bibnamefont
  {Sun}}, \ and\ \bibinfo {author} {\bibfnamefont {Arun}\ \bibnamefont
  {Bansil}},\ }\bibfield  {title} {\enquote {\bibinfo {title}
  {{Antiferromagnetic ground state of La$_2$CuO$_4$ : A parameter-free
  $ab-initio$ description}},}\ }\href {\doibase 10.1103/PhysRevB.98.125140}
  {\bibfield  {journal} {\bibinfo  {journal} {Physical Review B}\ }\textbf
  {\bibinfo {volume} {98}},\ \bibinfo {pages} {125140} (\bibinfo {year}
  {2018})}\BibitemShut {NoStop}%
\bibitem [{\citenamefont {Zhang}\ \emph
  {et~al.}(2020{\natexlab{b}})\citenamefont {Zhang}, \citenamefont {Furness},
  \citenamefont {Zhang}, \citenamefont {Wang}, \citenamefont {Zunger},\ and\
  \citenamefont {Sun}}]{Zhang2020b}%
  \BibitemOpen
  \bibfield  {author} {\bibinfo {author} {\bibfnamefont {Yubo}\ \bibnamefont
  {Zhang}}, \bibinfo {author} {\bibfnamefont {James}\ \bibnamefont {Furness}},
  \bibinfo {author} {\bibfnamefont {Ruiqi}\ \bibnamefont {Zhang}}, \bibinfo
  {author} {\bibfnamefont {Zhi}\ \bibnamefont {Wang}}, \bibinfo {author}
  {\bibfnamefont {Alex}\ \bibnamefont {Zunger}}, \ and\ \bibinfo {author}
  {\bibfnamefont {Jianwei}\ \bibnamefont {Sun}},\ }\bibfield  {title} {\enquote
  {\bibinfo {title} {Symmetry-breaking polymorphous descriptions for correlated
  materials without interelectronic u},}\ }\href {\doibase
  10.1103/PhysRevB.102.045112} {\bibfield  {journal} {\bibinfo  {journal}
  {Phys. Rev. B}\ }\textbf {\bibinfo {volume} {102}},\ \bibinfo {pages}
  {045112} (\bibinfo {year} {2020}{\natexlab{b}})}\BibitemShut {NoStop}%
\bibitem [{\citenamefont {Zhang}\ \emph {et~al.}(2022)\citenamefont {Zhang},
  \citenamefont {Singh}, \citenamefont {Lane}, \citenamefont {Kidd},
  \citenamefont {Zhang}, \citenamefont {Barbiellini}, \citenamefont
  {Markiewicz}, \citenamefont {Bansil},\ and\ \citenamefont
  {Sun}}]{ZhangR2020}%
  \BibitemOpen
  \bibfield  {author} {\bibinfo {author} {\bibfnamefont {Ruiqi}\ \bibnamefont
  {Zhang}}, \bibinfo {author} {\bibfnamefont {Bahadur}\ \bibnamefont {Singh}},
  \bibinfo {author} {\bibfnamefont {Christopher}\ \bibnamefont {Lane}},
  \bibinfo {author} {\bibfnamefont {Jamin}\ \bibnamefont {Kidd}}, \bibinfo
  {author} {\bibfnamefont {Yubo}\ \bibnamefont {Zhang}}, \bibinfo {author}
  {\bibfnamefont {Bernardo}\ \bibnamefont {Barbiellini}}, \bibinfo {author}
  {\bibfnamefont {Robert~S.}\ \bibnamefont {Markiewicz}}, \bibinfo {author}
  {\bibfnamefont {Arun}\ \bibnamefont {Bansil}}, \ and\ \bibinfo {author}
  {\bibfnamefont {Jianwei}\ \bibnamefont {Sun}},\ }\bibfield  {title} {\enquote
  {\bibinfo {title} {Critical role of magnetic moments in heavy-fermion
  materials: Revisiting ${\mathrm{smb}}_{6}$},}\ }\href {\doibase
  10.1103/PhysRevB.105.195134} {\bibfield  {journal} {\bibinfo  {journal}
  {Phys. Rev. B}\ }\textbf {\bibinfo {volume} {105}},\ \bibinfo {pages}
  {195134} (\bibinfo {year} {2022})}\BibitemShut {NoStop}%
\bibitem [{\citenamefont {Mostofi}\ \emph {et~al.}(2014)\citenamefont
  {Mostofi}, \citenamefont {Yates}, \citenamefont {Pizzi}, \citenamefont {Lee},
  \citenamefont {Souza}, \citenamefont {Vanderbilt},\ and\ \citenamefont
  {Marzari}}]{MOSTOFI20142309}%
  \BibitemOpen
  \bibfield  {author} {\bibinfo {author} {\bibfnamefont {Arash~A.}\
  \bibnamefont {Mostofi}}, \bibinfo {author} {\bibfnamefont {Jonathan~R.}\
  \bibnamefont {Yates}}, \bibinfo {author} {\bibfnamefont {Giovanni}\
  \bibnamefont {Pizzi}}, \bibinfo {author} {\bibfnamefont {Young-Su}\
  \bibnamefont {Lee}}, \bibinfo {author} {\bibfnamefont {Ivo}\ \bibnamefont
  {Souza}}, \bibinfo {author} {\bibfnamefont {David}\ \bibnamefont
  {Vanderbilt}}, \ and\ \bibinfo {author} {\bibfnamefont {Nicola}\ \bibnamefont
  {Marzari}},\ }\bibfield  {title} {\enquote {\bibinfo {title} {An updated
  version of wannier90: A tool for obtaining maximally-localised wannier
  functions},}\ }\href {\doibase https://doi.org/10.1016/j.cpc.2014.05.003}
  {\bibfield  {journal} {\bibinfo  {journal} {Computer Physics Communications}\
  }\textbf {\bibinfo {volume} {185}},\ \bibinfo {pages} {2309--2310} (\bibinfo
  {year} {2014})}\BibitemShut {NoStop}%
\bibitem [{\citenamefont {Guinea}\ \emph {et~al.}(1983)\citenamefont {Guinea},
  \citenamefont {Tejedor}, \citenamefont {Flores},\ and\ \citenamefont
  {Louis}}]{GreenF1983}%
  \BibitemOpen
  \bibfield  {author} {\bibinfo {author} {\bibfnamefont {F.}~\bibnamefont
  {Guinea}}, \bibinfo {author} {\bibfnamefont {C.}~\bibnamefont {Tejedor}},
  \bibinfo {author} {\bibfnamefont {F.}~\bibnamefont {Flores}}, \ and\ \bibinfo
  {author} {\bibfnamefont {E.}~\bibnamefont {Louis}},\ }\bibfield  {title}
  {\enquote {\bibinfo {title} {Effective two-dimensional hamiltonian at
  surfaces},}\ }\href {\doibase 10.1103/PhysRevB.28.4397} {\bibfield  {journal}
  {\bibinfo  {journal} {Phys. Rev. B}\ }\textbf {\bibinfo {volume} {28}},\
  \bibinfo {pages} {4397--4402} (\bibinfo {year} {1983})}\BibitemShut {NoStop}%
\bibitem [{\citenamefont {Sancho}\ \emph {et~al.}(1984)\citenamefont {Sancho},
  \citenamefont {Sancho},\ and\ \citenamefont {Rubio}}]{sancho1984quick}%
  \BibitemOpen
  \bibfield  {author} {\bibinfo {author} {\bibfnamefont {MP~Lopez}\
  \bibnamefont {Sancho}}, \bibinfo {author} {\bibfnamefont {JM~Lopez}\
  \bibnamefont {Sancho}}, \ and\ \bibinfo {author} {\bibfnamefont {Jessy}\
  \bibnamefont {Rubio}},\ }\bibfield  {title} {\enquote {\bibinfo {title}
  {Quick iterative scheme for the calculation of transfer matrices: application
  to mo (100)},}\ }\href
  {https://iopscience.iop.org/article/10.1088/0305-4608/14/5/016} {\bibfield
  {journal} {\bibinfo  {journal} {J. Phys. F: Met. Phys}\ }\textbf {\bibinfo
  {volume} {14}},\ \bibinfo {pages} {1205} (\bibinfo {year}
  {1984})}\BibitemShut {NoStop}%
\bibitem [{\citenamefont {Xu}\ \emph {et~al.}(2019)\citenamefont {Xu},
  \citenamefont {Song}, \citenamefont {Wang}, \citenamefont {Weng},\ and\
  \citenamefont {Dai}}]{Yuanfeng2019}%
  \BibitemOpen
  \bibfield  {author} {\bibinfo {author} {\bibfnamefont {Yuanfeng}\
  \bibnamefont {Xu}}, \bibinfo {author} {\bibfnamefont {Zhida}\ \bibnamefont
  {Song}}, \bibinfo {author} {\bibfnamefont {Zhijun}\ \bibnamefont {Wang}},
  \bibinfo {author} {\bibfnamefont {Hongming}\ \bibnamefont {Weng}}, \ and\
  \bibinfo {author} {\bibfnamefont {Xi}~\bibnamefont {Dai}},\ }\bibfield
  {title} {\enquote {\bibinfo {title} {Higher-order topology of the axion
  insulator ${\mathrm{euin}}_{2}{\mathrm{as}}_{2}$},}\ }\href {\doibase
  10.1103/PhysRevLett.122.256402} {\bibfield  {journal} {\bibinfo  {journal}
  {Phys. Rev. Lett.}\ }\textbf {\bibinfo {volume} {122}},\ \bibinfo {pages}
  {256402} (\bibinfo {year} {2019})}\BibitemShut {NoStop}%
\bibitem [{\citenamefont {Nogaki}\ \emph {et~al.}(2021)\citenamefont {Nogaki},
  \citenamefont {Daido}, \citenamefont {Ishizuka},\ and\ \citenamefont
  {Yanase}}]{Nogaki:PRR2021}%
  \BibitemOpen
  \bibfield  {author} {\bibinfo {author} {\bibfnamefont {Kosuke}\ \bibnamefont
  {Nogaki}}, \bibinfo {author} {\bibfnamefont {Akito}\ \bibnamefont {Daido}},
  \bibinfo {author} {\bibfnamefont {Jun}\ \bibnamefont {Ishizuka}}, \ and\
  \bibinfo {author} {\bibfnamefont {Youichi}\ \bibnamefont {Yanase}},\
  }\bibfield  {title} {\enquote {\bibinfo {title} {Topological crystalline
  superconductivity in locally noncentrosymmetric
  ${\mathrm{cerh}}_{2}{\mathrm{as}}_{2}$},}\ }\href {\doibase
  10.1103/PhysRevResearch.3.L032071} {\bibfield  {journal} {\bibinfo  {journal}
  {Phys. Rev. Res.}\ }\textbf {\bibinfo {volume} {3}},\ \bibinfo {pages}
  {L032071} (\bibinfo {year} {2021})}\BibitemShut {NoStop}%
\bibitem [{\citenamefont {Fu}\ and\ \citenamefont {Berg}(2010)}]{FuB2010}%
  \BibitemOpen
  \bibfield  {author} {\bibinfo {author} {\bibfnamefont {Liang}\ \bibnamefont
  {Fu}}\ and\ \bibinfo {author} {\bibfnamefont {Erez}\ \bibnamefont {Berg}},\
  }\bibfield  {title} {\enquote {\bibinfo {title} {Odd-parity topological
  superconductors: Theory and application to
  ${\mathrm{cu}}_{x}{\mathrm{bi}}_{2}{\mathrm{se}}_{3}$},}\ }\href {\doibase
  10.1103/PhysRevLett.105.097001} {\bibfield  {journal} {\bibinfo  {journal}
  {Phys. Rev. Lett.}\ }\textbf {\bibinfo {volume} {105}},\ \bibinfo {pages}
  {097001} (\bibinfo {year} {2010})}\BibitemShut {NoStop}%
\bibitem [{\citenamefont {Sato}(2010)}]{Sato:prb2010}%
  \BibitemOpen
  \bibfield  {author} {\bibinfo {author} {\bibfnamefont {Masatoshi}\
  \bibnamefont {Sato}},\ }\bibfield  {title} {\enquote {\bibinfo {title}
  {Topological odd-parity superconductors},}\ }\href {\doibase
  10.1103/PhysRevB.81.220504} {\bibfield  {journal} {\bibinfo  {journal} {Phys.
  Rev. B}\ }\textbf {\bibinfo {volume} {81}},\ \bibinfo {pages} {220504}
  (\bibinfo {year} {2010})}\BibitemShut {NoStop}%
\bibitem [{\citenamefont {Mong}\ \emph {et~al.}(2010)\citenamefont {Mong},
  \citenamefont {Essin},\ and\ \citenamefont {Moore}}]{Moore:PRB2010}%
  \BibitemOpen
  \bibfield  {author} {\bibinfo {author} {\bibfnamefont {Roger S.~K.}\
  \bibnamefont {Mong}}, \bibinfo {author} {\bibfnamefont {Andrew~M.}\
  \bibnamefont {Essin}}, \ and\ \bibinfo {author} {\bibfnamefont {Joel~E.}\
  \bibnamefont {Moore}},\ }\bibfield  {title} {\enquote {\bibinfo {title}
  {Antiferromagnetic topological insulators},}\ }\href {\doibase
  10.1103/PhysRevB.81.245209} {\bibfield  {journal} {\bibinfo  {journal} {Phys.
  Rev. B}\ }\textbf {\bibinfo {volume} {81}},\ \bibinfo {pages} {245209}
  (\bibinfo {year} {2010})}\BibitemShut {NoStop}%
\end{thebibliography}%



\end{document}